\documentclass[12pt,oneside,onecolumn,floatfix]{article}
\usepackage{epsfig}
\usepackage{amsmath,amssymb,amsfonts}
\usepackage{a4wide}
\usepackage[top=30pt,bottom=30pt,left=48pt,right=46pt]{geometry}
\usepackage{slashed, enumitem}
\usepackage[utf8]{inputenc}
\usepackage{jcappub}
\usepackage{enumerate}
\usepackage{float,color}
\usepackage{subfigure}
\usepackage{multirow}
\usepackage{slashed}
\usepackage{placeins}
\usepackage{wasysym} 
\usepackage{mathrsfs} 
\usepackage{caption}
\usepackage{amsmath}
\usepackage{amssymb}
\usepackage{graphicx}
\usepackage{slashed}
\usepackage{multirow}
\usepackage{placeins}
\usepackage[dvipsnames]{xcolor}
\usepackage{epstopdf}
\usepackage{soul}
\usepackage{tikz}
\usepackage[capitalise, english]{cleveref}
\usepackage{siunitx}
\usepackage{xspace}
\usepackage{booktabs}
\usetikzlibrary{trees}
\usetikzlibrary{decorations.pathmorphing}
\usetikzlibrary{decorations.markings}
\usepackage[colorlinks=true,citecolor=blue,linkcolor=blue]{hyperref}

\definecolor{americanrose}{rgb}{1.0, 0.01, 0.24}
\definecolor{ao}{rgb}{0.0, 0.0, 1.0}

\newcommand\myshade{80}
\colorlet{mylinkcolor}{violet}
\colorlet{mycitecolor}{americanrose}
\colorlet{myurlcolor}{ao}

\hypersetup{
	linkcolor  = mylinkcolor!\myshade!black,
	citecolor  = mycitecolor!\myshade!black,
	urlcolor   = myurlcolor!\myshade!black,
	colorlinks = true
}

\usepackage{array}
\newcolumntype{L}[1]{>{\raggedright\let\newline\\\arraybackslash\hspace{0pt}}m{#1}}
\newcolumntype{C}[1]{>{\centering\let\newline\\\arraybackslash\hspace{0pt}}m{#1}}
\newcolumntype{R}[1]{>{\raggedleft\let\newline\\\arraybackslash\hspace{0pt}}m{#1}}

\def \beq{\begin{equation}}
\def \eeq{\end{equation}}
\def \bea{\begin{eqnarray}}
\def \eea{\end{eqnarray}}
\def \ba{\begin{array}}
	\def \ea{\end{array}}

\title{Dark matter capture in celestial objects: light mediators, self-interactions, and complementarity with direct detection}

\author[a]{Basudeb Dasgupta,}
\emailAdd{bdasgupta@theory.tifr.res.in}
\affiliation[a]{Tata Institute of Fundamental Research,\\ Homi Bhabha
	Road, Mumbai 400005, India}

\author[b]{Aritra Gupta,} 
\emailAdd{ aritra.gupta@ulb.be} 
\affiliation[b]{Service de Physique Th\'eorique, Universit\'e Libre de Bruxelles,\\	Boulevard du Triomphe, CP225, 1050 Brussels, Belgium}

\author[a]{and Anupam Ray\,}
\emailAdd{anupam.ray@theory.tifr.res.in}

\abstract
{We generalize the formalism for DM capture in celestial bodies to account for arbitrary mediator mass, and update the existing and projected astrophysical constraints on DM-nucleon scattering cross section from observations of neutron stars. We show that the astrophysical constraints on the DM-nucleon interaction strength, that were thought to be the most stringent, drastically weaken for light mediators and can be completely voided. For asymmetric DM, existing astrophysical constraints are completely washed out for mediators lighter than 5\,MeV, and for annihilating DM the projected constraints are washed out for mediators lighter than 0.25\,MeV. Related terrestrial direct detection bounds also weaken, but in a complementary fashion; they supersede the astrophysical capture bounds for small or large DM mass, respectively for asymmetric or annihilating DM. Repulsive self-interactions of DM have an insignificant impact on the total capture rate, but a significant impact on the black hole formation criterion. This further weakens the constraints on DM-nucleon interaction strength for asymmetric self-repelling DM, whereas constraints remain unaltered for annihilating self-repelling DM.  We use the correct Hawking evaporation rate of the newly formed black hole, that was approximated as a blackbody in previous studies, and show that, despite a more extensive alleviation of collapse as a result, the observation of a neutron star collapse can probe a wide range of DM self-interaction strengths.}

\subheader{TIFR/TH/20-18 \newline ULB-TH/20-07}


\begin{document}
\maketitle

\section{Introduction}
\label{sec:intro}
Recent precise measurements of the cosmic microwave background anisotropies by Planck~\cite{Aghanim:2018eyx} has confirmed with a high degree of certainty about the existence of cold dark matter as a dominant component of the universe. However, very little is known about the fundamental properties of such dark matter (DM) particles, e.g., their mass, spin, and their strength of interaction with baryonic matter, as well as amongst themselves. A theoretically well-motivated candidate for dark matter is a Weakly-Interacting-Massive-Particle (WIMP). In a large class of models, a WIMP with mass ranging in the sub-GeV to several-hundred-TeV range can suitably freeze-out to satisfy the relic density of dark matter observed today~\cite{Jungman:1995df,Steigman:2012nb}. 
Thus, in order to directly observe such WIMP-like particles in nature, several terrestrial experiments have been set up. Despite decades of dedicated efforts to hunt these elusive particles, the experiments have only yielded null results. Therefore, non-observation of dark matter signals from these searches have severely constrained the allowed masses and interaction strengths of such particles~\cite{Aprile:2018dbl,Akerib:2016vxi,Cui:2017nnn,Agnes:2018oej,Ren:2018gyx,Abdelhameed:2019hmk,Armengaud:2019kfj,Abramoff:2019dfb,Agnese:2018col}. Apart from these terrestrial searches, several complementary cosmological and astrophysical probes have yet to reveal the nature of dark matter particles~\cite{Chivukula:1989cc,Raffelt:1999tx,Cyburt:2002uw,Fayet:2006sa,Mack:2007xj,Dvorkin:2013cea,Ali-Haimoud:2015pwa,Gluscevic:2017ywp,Raj:2017wrv,Slatyer:2018aqg,Boddy:2018kfv,Bhoonah:2018wmw,Xu:2018efh,Cappiello:2018hsu,Bringmann:2018cvk,Krnjaic:2019dzc,Gluscevic:2019yal,Wadekar:2019xnf,Nadler:2019zrb}. Nevertheless, these WIMP-like models remain the best motivated and most eminently testable, and yet incompletely tested.

Prominent among astrophysical searches, are those for anomalous signatures of DM particles captured inside celestial objects. DM particles from the surrounding halo can collide with the stellar constituents, lose a sufficient amount of energy, and get trapped inside the stellar core after one or more scatterings~\cite{Press:1985ug,Gould:1987ju,Gould:1987ir, Bramante:2017xlb, Dasgupta:2019juq}. WIMP-like DM particles, that are purely asymmetric, do not annihilate away due to the dearth of antiparticles, allowing a runaway accumulation inside the stellar body. Gradual accretion of such asymmetric DM particles inside a stellar object can result in the formation of a black hole and in due time the newly formed black hole can even destroy the stellar object. Upper limits on the DM-nucleon scattering cross section can thus be obtained from the mere existence of such objects today~\cite{Goldman:1989nd,Gould:1989gw,Bertone:2007ae,deLavallaz:2010wp,Kouvaris:2010jy,McDermott:2011jp,Kouvaris:2011fi,Guver:2012ba,Kouvaris:2012dz,Kouvaris:2013kra,Bell:2013xk,Bramante:2013hn,Jamison:2013yya,Bramante:2013nma,Bramante:2014zca,Garani:2018kkd,Lin:2020zmm}. Alternatively, the collapse of a black hole may spark a supernova, resulting in the destruction of the star, which is also constrained~\cite{Bramante:2015cua,Acevedo:2019gre,Graham:2018efk,Janish:2019nkk}. On the other hand, dark matter particles that are allowed to annihilate among themselves can lead to heating of the stellar core, thereby, affecting its observed luminosity. Apart from this, DM particles can also kinetically heat up a celestial object by depositing the excess kinetic energy gained during its gravitational capture. Such complementary limits on DM-nucleon interaction strengths have been derived from observations of cold neutron stars~\cite{Kouvaris:2007ay,Kouvaris:2010vv,deLavallaz:2010wp,Bramante:2017xlb,Raj:2017wrv,Baryakhtar:2017dbj,Bell:2018pkk,Chen:2018ohx,Acevedo:2019agu} and white dwarfs~\cite{Bertone:2007ae,McCullough:2010ai,Hooper:2010es,Dasgupta:2019juq}. However, owing at least partly to the lack of an appropriate formalism, all these studies have only considered DM-nucleon scattering mediated by a relatively massive mediator.

In addition to the searches for these WIMP-like particles, there are a few anomalies associated with the small scale structures of the universe, that may require new physics beyond the standard $\Lambda$CDM cosmology. A widely discussed solution to these small scale anomalies is to introduce strong self-interactions among the DM particles mediated by lighter mediators~\cite{Spergel:1999mh,Dave:2000ar,Loeb:2010gj,Aarssen:2012fx,Dasgupta:2013zpn,Chu:2014lja,Kaplinghat:2015aga,Kamada:2016euw,Tulin:2017ara,Bullock:2017xww}, roughly at the level of $\sigma_{\chi \chi}/m_{\chi} \sim (0.1-1)\,\rm cm^{2} \,\rm /g$. DM self scattering cross section per unit mass, $\sigma_{\chi \chi}/m_{\chi}$, is however constrained, e.g., from the measurement of the size of offset between the DM halo and that of the baryonic matter in colliding galaxy clusters \cite{Dawson:2011kf,Merten:2011wj,Bradac:2008eu,Randall:2007ph,Robertson:2016xjh}. The most frequently quoted among these upper limits is the one obtained from the Bullet Cluster which states  $\sigma_{\chi \chi}/m_{\chi} \lesssim 1\,\rm cm^{2} \,\rm /g$. These two requirements are in tension, that may be resolved if one appeals to a light mediator that allows a velocity dependent cross section that is large in galaxies but small for cluster collisions.

Astrophysical probes such as gravitational collapse of captured DM particles inside a stellar object can also be used to determine DM self-interaction strength. Given nonzero self-interaction among the DM particles, an incoming dark matter particle can also collide with previously captured dark matter particles inside the stellar core. This additional contribution is known as the self-capture of DM particles. The accretion of DM particles with non negligible self-interaction and its related phenomenology is extensively studied in~\cite{Zentner:2009is,Kouvaris:2011fi,Kouvaris:2011gb,Guver:2012ba,Fan:2013bea,Chen:2015uha,Chen:2015bwa,Feng:2016ijc,Chen:2018ohx,Gaidau:2018yws}.  For gravitational collapse, the total number of captured DM particles has to exceed the number of DM particles required for black hole formation. For strong repulsive self-interactions, the number of dark matter particles required for a successful collapse of the star is quite large. On the other hand, due to the enormous baryonic density of compact objects, the contribution to the total number of captured dark matter particles from self-capture is negligible as compared to that from the baryonic capture. Hence, it is not possible to constrain strongly self interacting dark matters using such accretion scenarios. However one can probe very feeble self-interactions using gravitational collapse of captured dark matter particle and thereby place constraints which are complementary to those obtained from the observations of colliding galaxies. Thus, although DM self-interactions best motivated with a light mediator, such light mediators have not been consistently included in previous studies.

In this paper, we calculate the capture rate of dark matter particles that are gravitationally attracted towards a stellar object for interactions mediated by light mediators. A key quantity in the calculation is the dependence of the fractional loss in kinetic energy of the incoming particle, with the scattering angle. We find that it is intimately related to the differential cross section of DM-target scattering. The distribution of the scattering angle is uniform only for isotropic differential cross section. This happens when the mass of the particle mediating the interaction is much larger than the corresponding momentum transfer. However, in general, the differential cross section may involve lighter mediators and the distribution is not guaranteed to be uniform. We use the general form of energy loss distribution to estimate the baryonic and self-capture rate for DM interactions mediated by a Yukawa potential. We apply our general treatment to update the existing and projected astrophysical constraints on DM-nucleon scattering cross section for both annihilating and asymmetric DM. We find that these constraints depend sensitively on mediator mass, weaken appreciably for light mediators, and can be completely washed out for sufficiently light mediators. We show that inclusion of repulsive self-interactions among the DM particles further enfeeble the constraints on DM-nucleon interaction strength for asymmetric DM. However, for annihilating DM, constraints obtained from dark heating, remain unaltered with inclusion of DM  self-interactions due to its insignificant contribution to the total capture rate. We also show that gravitational collapse of captured DM particles acts as an astrophysical probe of DM self scattering cross section and the constraints from gravitational collapse are complementary to the upper limits obtained from dwarf galaxies and galaxy clusters.

The paper is structured as follows: In Sec.\,\ref{sec:formalism}, we have generalized the treatment of DM capture in celestial objects for arbitrary mediator mass, followed by Sec.\,\ref{sec:results} discussing about our results. We summarize and conclude in Sec.\,\ref{sec:summ}.

\section{DM capture for arbitrary mediator mass}
\label{sec:formalism}
\subsection{Derivation of energy loss distribution}

\begin{figure}
\begin{center}
	\includegraphics[width=0.8\textwidth]{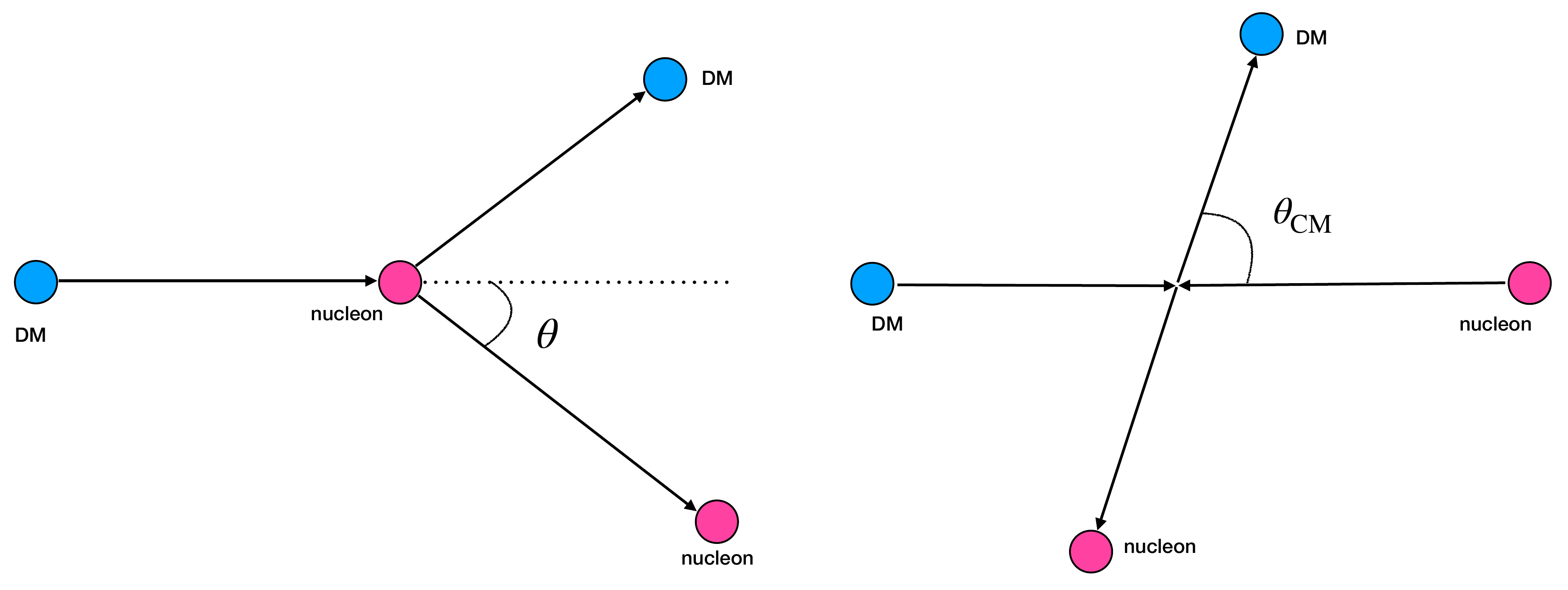}
\end{center}	
	\caption{ DM-nucleon scattering is schematically illustrated in the rest frame of nucleon (left) as well as in the center of mass frame (right). The recoil angle in the rest frame of nucleon is denoted as $\theta$ and the scattering angle in the center of mass frame as $\theta_{\rm CM}$. The probability distribution function of either of these angles determines the energy loss distribution.}
	\label{fig:labcom}
\end{figure}

For DM-nucleon interaction via a Yukawa potential,
\begin{equation}
V=\frac{\alpha}{r}e^{-m_{\phi}r}\,,
\end{equation}
where $\alpha$ parametrizes the interaction strength and $m_{\phi}$ denotes the mediator mass, the  differential scattering cross section in non-relativistic Born approximation is given by
\begin{equation}
	\frac{d\sigma}{d\Omega_{ \rm CM}} = \,\frac{4\mu^2 \alpha^2}{\left(4\mu^2v^2_{\rm rel}\sin^2(\theta_{\rm CM}/2)+m^2_{\phi} \right)^2}\,.
\end{equation}
Here $\mu$ is the reduced mass of the system, $v_{\rm rel}$ is the relative velocity between DM and nucleon made dimensionless in units of the speed of light, and $\theta_{\rm CM}$ denotes the scattering angle in the center of mass frame. The fractional loss in kinetic energy in DM-nucleon scattering, as shown in Fig.\,\ref{fig:labcom}, is given by 
\begin{equation}
\frac{\Delta E}{E}= \frac{4m_{\chi}m_n}{(m_{\chi}+m_n)^2}z=\beta z\,,
\end{equation}
where $m_{\chi}$, $m_n$ are the mass of DM and nucleon respectively. The scattering kinematics determines  $z = \cos^2 \theta = \sin^2 (\theta_{\rm CM}/2)$, with $z \in $ [0,1]. 

\begin{figure}
 	\centering
 	\includegraphics[scale=0.55]{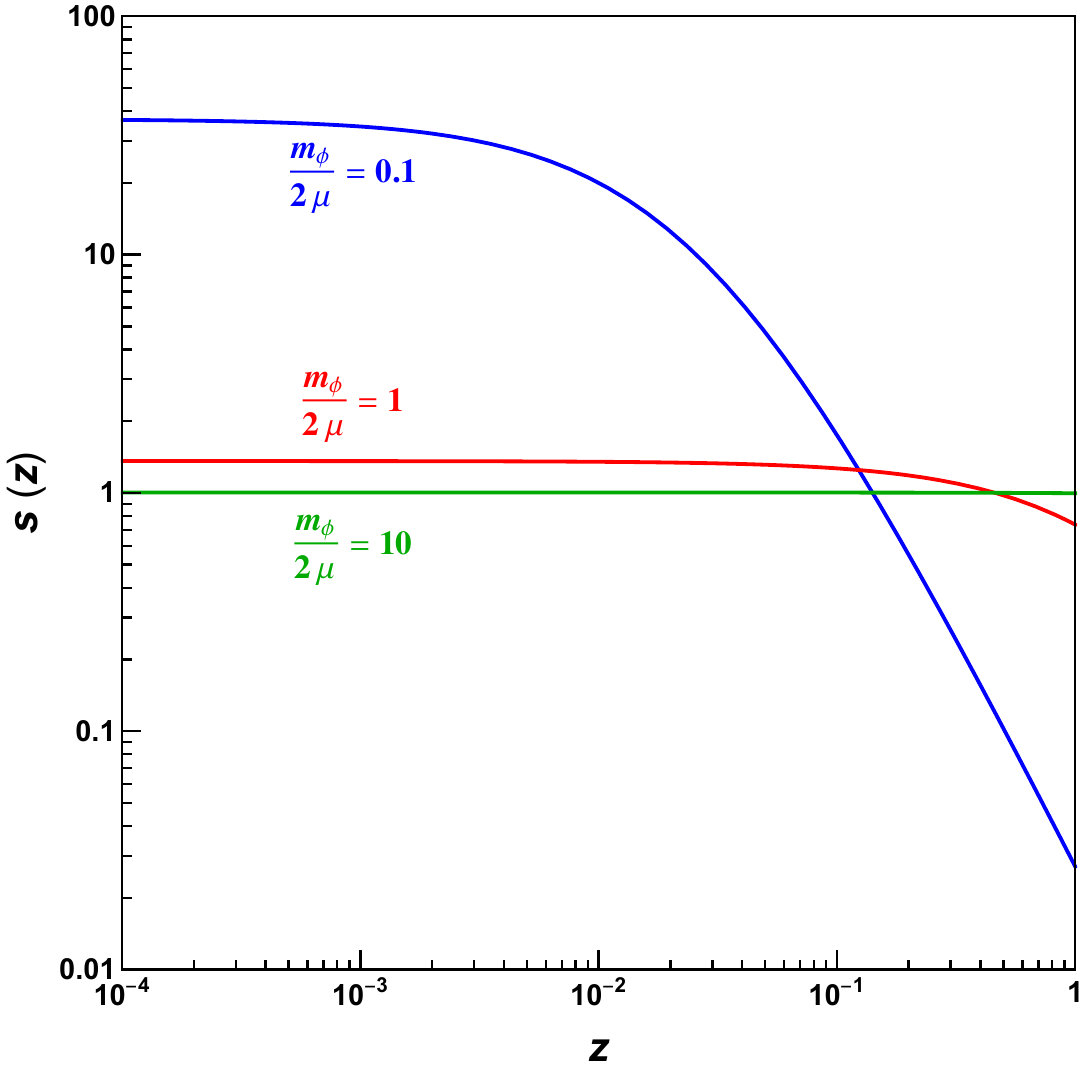}
 	\caption{Energy loss distribution, $s(z)$, with $z$ for three representative values of $m_{\phi}/2\mu$. When the mediator is lighter than the reduced mass, the deviation from uniform energy loss distribution becomes significant.}
 	\label{fig:gen}	
 \end{figure}

The \emph{energy loss distribution} $s(z)$, a key quantity to estimate the capture rate precisely, is determined by the distribution of $\Omega_{\rm CM}$, which is in turn dictated by the differential scattering cross section  of the relevant scattering process,
\begin{equation}
s(\Omega_{ \rm CM})=\frac{1}{\sigma}\frac{d\sigma}{d\Omega
	_{ \rm CM}}=\frac{1}{4\pi}s(z)\,.
\end{equation}
For DM-nucleon scattering via a Yukawa potential, to consider a widely applicable example, the energy loss distribution $s(z)$ is given by
\begin{equation}\label{eloss}
s(z)=\frac{m^2_{\phi}\left(4\mu^2v^2_{\rm rel}+m^2_{\phi} \right)}{\left(4\mu^2v^2_{\rm rel}z+m^2_{\phi} \right)^2}\,.
\end{equation}
For DM self scattering via a Yukawa potential, the reduced mass $\mu=m_{\chi}/2$, and energy loss distribution $s^{\textrm{self}}(z)$ simplifies to
\begin{equation}\label{self}
s^{\textrm{self}}(z)=\frac{m^2_{\phi}\left(m_{\chi}^2v^2_{\rm rel}+m^2_{\phi} \right)}{\left(m_{\chi}^2v^2_{\rm rel}z+m^2_{\phi} \right)^2}\,.
\end{equation}
In the limit $m_{\phi} \to \infty$, we recover the familiar expressions for uniform distribution, $s(z)=1 \,\textrm{and} \,s^{\textrm{self}}(z)=1$, which have been used extensively in the previous treatments. Variation of $s(z)$ with mediator mass is shown in Fig.\,\ref{fig:gen} for DM-nucleon interaction mediated via a Yukawa potential. From Fig.\,\ref{fig:gen}  it is evident that the assumption of uniformity in energy loss distribution, i.e., $s(z)=1$, is a poor approximation when the mediator is lighter than either of the scattering particles. 
 
\subsection{Analytical formula for the capture rate}
Far away from the stellar body, the dark matter particle has a velocity $u$ and when it reaches the surface of the stellar body, its velocity becomes  $w$, given by
\begin{equation}
w^2 = u^2 + v_{\rm esc}^2\,.
\end{equation}
It undergoes one or more scatterings as it transits through the stellar
object. As a result of these collisions with essentially static stellar constituents, the incoming dark matter particle can lose energy. If eventually its final velocity $v_{\rm f}$ becomes less than the escape velocity  $v_{\rm esc}$ of the stellar object, it is considered to be captured. Capture of a dark matter particle can occur via a single or multiple scatterings with stellar constituents~\cite{Bramante:2017xlb,Dasgupta:2019juq,Ilie:2020vec}. In this work, we assume that the dark matter particle is captured after a single scattering. This is reasonable, because, for dark matter particles which are lighter than a few 100 TeV, capture almost always occurs after a single scattering~\cite{Dasgupta:2019juq}. Once the DM particle gets captured, it loses more and more energy via scattering with the nucleons. As a consequence its velocity keeps decreasing, until it attains thermal equilibrium with the stellar constituents and joins an isothermal sphere of radius $r_{\rm th}$ around the core. This process is known as \textit{thermalization} of captured DM particles and the thermalization timescale depends solely on the DM-nucleon scattering cross section.

The capture rate of dark matter particles in the stellar body depends on the number of stellar constituents $N_\textrm{n}$, scattering cross section $\sigma_{\chi \textrm{n}}$, the flux of incoming dark matter particles, and most importantly on $g_1(u)$, the probability of incurring energy loss  in a single collision. Therefore, the \emph{capture rate} takes the form
\begin{equation}\label{capt}
C =\frac{\rho_{\chi}}{m_{\chi}} \int \dfrac{f(u)du}{u}\,(u^2+v_{\rm esc}^2)\,N_\textrm{n}\, \textrm{Min} \left[\sigma_{\chi \textrm{n}},\sigma_{\chi \textrm{n}}^{\rm sat}\right] \, g_1(u) \,,
\end{equation}
where $\sigma_{\chi \textrm{n}}^{\rm sat}$ denotes the geometrical saturation limit of DM-nucleon scattering cross section,
\begin{equation}
\sigma_{\chi \textrm{n}}^{\rm sat}= \pi R^2/N_\textrm{n}\,.
\end{equation}
In the case of DM scattering with nucleons, the effect of Pauli blocking must be taken into account~\cite{McDermott:2011jp}. Due to Pauli blocking, if the momentum transfer $\Delta p$ is less than the Fermi momentum $p_F$, all the nucleons in the stellar body are not available to scatter with incoming DM particles. Therefore, the total number of targets $N_\textrm{n}$ and hence the capture rate should be suppressed by a factor of Pauli blocking efficiency $\zeta$ which is approximated by \rm Min $\left[\Delta p/p_F, 1\right]$. See~\cite{Garani:2018kkd} for an improved treatment of Pauli blocking efficiency. However, for a neutron star, due to its enormous nucleon density and very high escape velocity, typical momentum transfer is always greater than Fermi momentum  for DM mass above  1 GeV. Therefore, for DM mass above  1 GeV, Pauli blocking efficiency $\zeta$ becomes unity for DM capture in neutron stars.

In order to estimate the capture rate in Eq.\,(\ref{capt}), the probability of incurring energy loss $ g_1(u)$, which is often quoted as \textit{capture probability}, has to be computed. It depends on the energy loss distribution, $s(z)$, and  is given by
\begin{equation}\label{g1def}
g_1(u)= \int_{0}^{1} dz \,\Theta \left(v_{\rm esc}- v_{\rm f} \right)s(z)\,,
\end{equation}
where $v_{\rm f} = \sqrt{(u^2+v^2_{\rm esc}) (1-z\beta)} $ denotes the final velocity of the DM particle.
Using the analytical expression for $s(z)$ from Eq.\,(\ref{eloss}), capture probability for DM-nucleon interaction mediated via a Yukawa potential simplifies to
\begin{equation}\label{capture}
g_1(u)= \frac{m^2_{\phi} \left(1-\frac{1}{\beta}\frac{u^2}{u^2+v^2_{\rm esc}}\right)}{\left(m^2_{\phi}+\frac{4\mu^2u^2}{\beta c^2}\right)} \,\Theta \left(v_{\rm esc}\sqrt{\frac{\beta}{1-\beta}}-u\right)\,.
\end{equation}
Note that, in the limit of $m_{\phi} \to \infty$, we again recover the familiar expression for capture probability~\cite{Gould:1987ir}. From this generalized analytical expression of the capture probability, one can easily estimate the capture rate of DM particles inside a stellar body by considering a suitable velocity distribution of dark matter particles. See~\cite{Joglekar:2020liw,Bell:2020jou} for a recent treatment of capture rate of DM particles in neutron stars in the contact interaction approximation.

\subsection{Analytical formula for the self-capture rate}

In the previous section, we have estimated the baryonic capture rate for DM-nucleon scattering via a Yukawa potential. If the dark matter particles have appreciable self-interaction strength, an incoming dark matter particle can also lose its energy by colliding with  previously captured DM particles within the celestial core. This is known as \textit{self-capture} of DM particles. In this section, we estimate the generalized self-capture rate for DM self-interactions mediated by a Yukawa potential.

For self-capture, the incoming dark matter particle has to lose enough energy so that its final velocity $v_{\rm f}$ falls below the escape velocity $v_{\rm esc}$ of the stellar body. The target dark matter particle, on the other hand,  gains energy from these collisions but its final velocity $v^{\prime}_{\rm f}$ should also remain less than $v_{\rm esc}$. Therefore, the \textit{self-capture probability} is given by
\begin{equation}
g^{\rm self}_1(u)= \int_{0}^{1} dz\, \Theta(v_{\rm esc}-v_{\rm f}) \, \Theta(v_{\rm esc}-v^{\prime}_{\rm f})\,s^{\textrm{self}}(z)\,,
\end{equation}
where $v_{\rm f} = \sqrt{(u^2+v^2_{\rm esc})(1-z)}$ denotes the final velocity of the incoming DM particle  and $v^{\prime}_{\rm f} = \sqrt{(u^2+v^2_{\rm esc})z}$ denotes the final velocity of the target DM particle. 

Using the expression for $s^{\textrm{self}}(z)$ from Eq.\,(\ref{self}), the generalized self-capture probability for DM self-interactions mediated by a Yukawa potential simplifies to
\begin{equation}\label{pp}
g^{\rm self}_1(u)=\frac{m^2_{\phi}\left(m^2_{\phi}+m^2_{\chi}\frac{u^2+v^2_{\rm esc}}{c^2}\right)}{\left(m^2_{\phi}+m^2_{\chi}\frac{v^2_{\rm esc}}{c^2}\right)\left(m^2_{\phi}+m^2_{\chi}\frac{u^2}{c^2}\right)} \left(\frac{v^2_{\rm esc}-u^2}{v^2_{\rm esc}+u^2} \right)\,\Theta(v_{\rm esc}-u)\,.
\end{equation}
Note that, in the limit of $m_{\phi} \to \infty$, we recover the familiar expression for {self-capture probability}~\cite{Zentner:2009is}.

Self-capture rate of dark matter particles  in the stellar body depends on the number of already captured DM particle ${N}_{\chi}$,  DM self scattering cross section $\sigma_{\chi \chi}$, the flux of incoming dark matter particles and most importantly on the probability of incurring energy loss $g^{\rm self}_1(u)$ which depends on the energy loss distribution. Hence, the self-capture rate takes the form
\begin{equation}
C^{\textrm {self}}=\frac{\rho_{\chi}}{m_{\chi}} \int \dfrac{f(u)du}{u}\,(u^2+v_{\rm esc}^2) \, N_{\chi}\, \textrm {Min}\, \left[\sigma_{\chi \chi},\sigma^{\rm sat}_{\chi \chi}\right]\,g^{\rm self}_1(u)\,.
\end{equation}

Using the analytical expression of the self-capture probability from Eq.\,(\ref{pp}), one can easily estimate the generalized self-capture rate  for a given velocity distribution of dark matter particles. The self-capture rate depends linearly on the number of captured dark matter, ${N}_{\chi}$, so the number of captured particle due to self-capture grows exponentially with time. In general, the equation governing the number of captured DM in presence of self-interaction is given by
\begin{equation}
\dfrac{dN_\chi}{dt}=C+C^{\rm self}.
\end{equation}
However, this exponential growth due to self-capture cuts off when DM self-interaction strength reaches its geometric saturation value $\sigma^{\rm sat}_{\chi \chi}$~\cite{Guver:2012ba,Chen:2018ohx}, determined by the radius of the thermalization sphere $r_{\rm th}$ inside the stellar body
\begin{equation}
N_{\chi}\sigma^{\rm sat}_{\chi \chi}= \pi r_{\rm th}^2\,.
\end{equation}
Once the number of captured dark matter particles inside the stellar object attains its saturation value, self-capture rate becomes time-independent, i.e., $N_{\chi}\sigma_{\chi \chi} \to \pi r_{\rm th}^2$. Therefore, after the saturation, number of captured dark matter particles due to self-interactions grows only linearly with time.
Note that, for self-capture of fermionic DM, the effect of Pauli blocking has to be taken into consideration and hence the self-capture rate of fermionic DM is additionally suppressed by Pauli blocking efficiency $\zeta$. However, in case of bosonic DM, there is no such suppression in self-capture rate.

\section{Constraining DM interactions using neutron stars}
\label{sec:results}

We compute the capture rate of DM particles, inside a neutron star by considering the velocity distribution of incoming DM particles as Maxwell-Boltzmann with halo velocity dispersion  220 km/s. We take typical values for neutron star parameters~\cite{McDermott:2011jp}, mass $M_{\rm NS} = 1.44 M_{\odot}$, radius $R_{\rm NS} = 10.6$ km and the central density $\rho_{\rm NS} = 1.4 \times 10^{15}$ g/$\rm cm^3$ in this work. We have not considered the effect of multiple scatterings in capture rate, because, for DM particles lighter than a few 100 TeV, capture almost always occurs after a single scattering~\cite{Dasgupta:2019juq}. 
 
While considering DM capture rate in a neutron star, two corrections, gravitational blueshift of the initial kinetic energy and general relativistic enhancement of the number of dark matter particles crossing the stellar surface, have to be considered~\cite{Bramante:2017xlb}. For the gravitational blueshift effect, in the rest frame of the neutron star, incoming DM particle has a relativistic correction in kinetic energy $\sim \mathcal{O} (v^2_{\rm esc}/c^2)$, and it modifies the capture probability. As the lower limit of the capture probability integral is proportional to $v^2_{\rm esc}/w^2$, this can be accounted for by substituting
\begin{equation}
v'_{\rm esc} \to \frac{v_{\rm esc}}{\left(1+\frac{3}{4} \left(\frac{v_{\rm esc}}{c}\right)^2\right)^{1/2}}\,.
\end{equation}
For the general relativistic enhancement of the gravitational potential of the stellar body, the number of dark matter particles which traverse the stellar surface increases. Therefore, this effect modifies the capture rate by~\cite{Kouvaris:2007ay,Bramante:2017xlb}
\begin{equation}
C \to \frac{C}{1- \left(\frac{v'_{\rm esc}}{c}\right)^2}\,.
\end{equation}
We have explicitly examined that the gravitational blueshift effect can reduce the capture rate up to $\sim 20 \%$, whereas, the general relativistic correction can enhance the capture rate by as far as a factor of two.

\begin{figure}
	\centering
	\includegraphics[angle=0.0,width=0.45\textwidth]{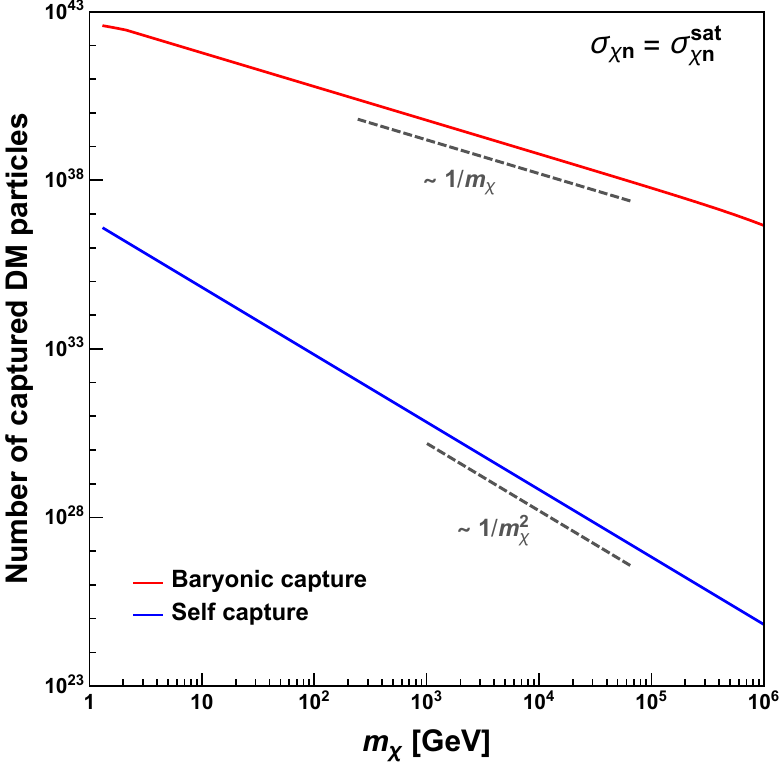}\quad\quad
	\includegraphics[angle=0.0,width=0.45\textwidth]{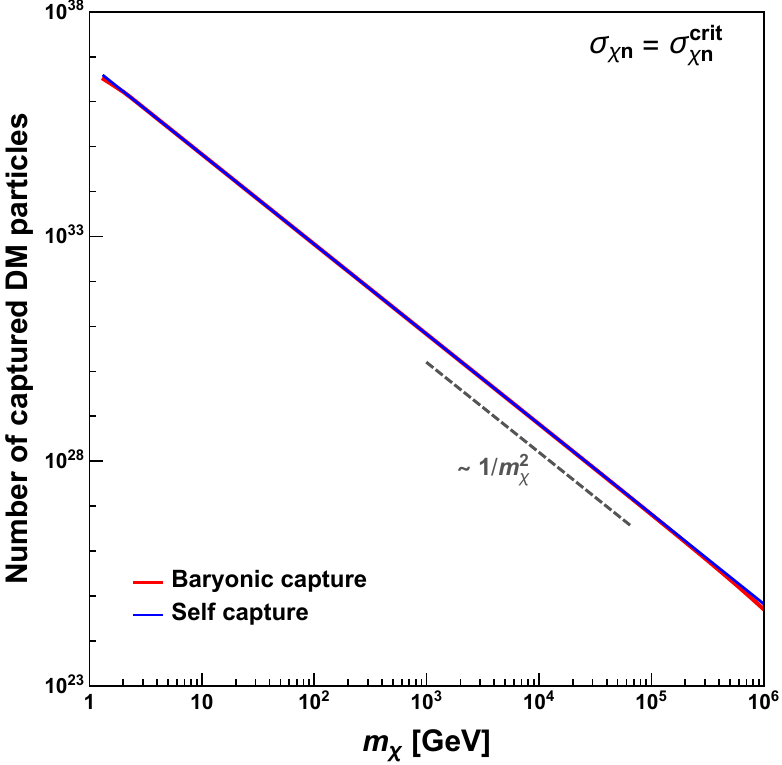}
	\caption{Number of DM particles captured inside a neutron star is shown with DM mass, assuming uniform energy loss distribution. In the left panel, DM-nucleon scattering cross section is taken to its largest possible value, i.e., the geometrical saturation value, $2\times 10^{-45} \, \rm cm^2$. In the right panel, the scattering cross section is taken to its critical value, $\pi r_{\rm th}^2/N_\textrm{n}$, such that self-capture and baryonic capture are equally efficient. In each plot, the red lines correspond to baryonic capture and the blue lines correspond to the maximum allowed self-capture. The local DM density around the neutron star is taken as 0.4 GeV/$\rm cm^3$ and age of the neutron star is taken as 6.69$\, \times 10^9$ years. The core temperature of the neutron star is taken as $2.1\times 10^6\,$K and all the other neutron star parameters are described in the text. We restrict our study to $m_{\chi}<10^{6}$\,GeV to ensure multiple scattering is not relevant.}
	\label{fig:comp}	
\end{figure}

We have also neglected the contribution from self-capture in the estimation of the total capture rate. Self-interactions among the DM particles can lead to an additional contribution to the capture rate. For the not too small values of $\sigma_{\chi\textrm{n}}$ that we can probe, baryonic capture always dominates over the self-capture even if we take the maximum allowed value of the DM self-interaction strength. This can be understood easily, because in neutron stars, the extremely large nucleon density  increases the baryonic capture rate and also shrinks the thermalization sphere as $r_{\rm th}=\sqrt{9k_{\rm B}T_{\rm NS}/4 \pi G\rho_{\rm NS}m_{\chi}}$. Since the self-capture rate scales quadratically with the radius of the thermalization sphere $r_{\rm th}$, self-capture rate falls off significantly. However, if the DM-nucleon scattering cross section is extremely small, at some point, self-capture  dominates over the baryonic capture. This is what we call the \emph{critical cross section}, $\sigma_{\chi \textrm{n}}^{\rm crit}$, which is given by $\sigma_{\chi \textrm{n}}^{\rm crit} \sim \pi r_{\rm th}^2/N_\textrm{n}$. Quantitatively, we can estimate that for a neutron star with core temperature of $2.1\times10^6\,$K and for DM mass of 100 GeV,  this limiting cross section turns out to be $\sim 2.5\times 10^{-53} \ \rm cm^2$. For heavier dark matter, this limiting cross section further reduces linearly with DM mass, as $r_{\rm th}^2$ is inversely proportional to the DM mass. Of course, thermalization requires $\sigma_{\chi \textrm{n}}$ to not become smaller than the minimum value required for thermalization. For $m_{\chi}\gtrsim1$\,TeV, the critical cross section falls below the minimum thermalization cross section, and self-capture never exceeds baryonic capture for even moderately large DM masses.

Total number of captured DM particles due to baryonic and self-interactions is shown in Fig.\,\ref{fig:comp}, for a neutron star with a core temperature of $2.1\times10^6\,$K. From Fig.\,\ref{fig:comp}, it is evident that self-capture can contribute significantly to the total capture rate only when the DM-nucleon interaction strength is lowered past its critical value which is really small compared to the parameter space of contemporary interest. Note that baryonic capture scales as $1/m_{\chi}$, as the number density of incoming DM particles is inversely proportional to the DM mass. In the right panel, the additional $1/m_{\chi}$ suppression in the baryonic capture simply comes from the mass dependence of DM-nucleon scattering cross section, as $\sigma_{\chi \textrm{n}}^{\rm crit}  \sim 1/m_{\chi}$. However, number of captured DM particles due to self-capture always scales as $\sim 1/m^2_{\chi}$, because self-capture rate is quadratically proportional to the thermalization radius, and $r^2_{\rm th}$ scales inversely to the DM mass, in addition to the $1/m_{\chi}$ suppression from the number density of the incoming DM particles.

\subsection{Annihilation of captured DM \textit{\&} kinetic heating}

For annihilating DM, the number of captured dark matter particles inside the stellar object follows
\begin{equation}
\frac{dN_{\chi}}{dt} = C-C_a N_{\chi}^2\,,
\end{equation}
where, $C$ is the total capture rate and $C_a$ is the annihilation rate. The annihilation rate is simply given by $C_a = 3\langle \sigma_a v \rangle/ 4\pi r^3_{\rm th}$, where $\langle \sigma_a v \rangle$ denotes the thermally averaged annihilation cross section of the DM particles.

\begin{figure}
\begin{center}
	\includegraphics[angle=0.0,width=0.8\textwidth]{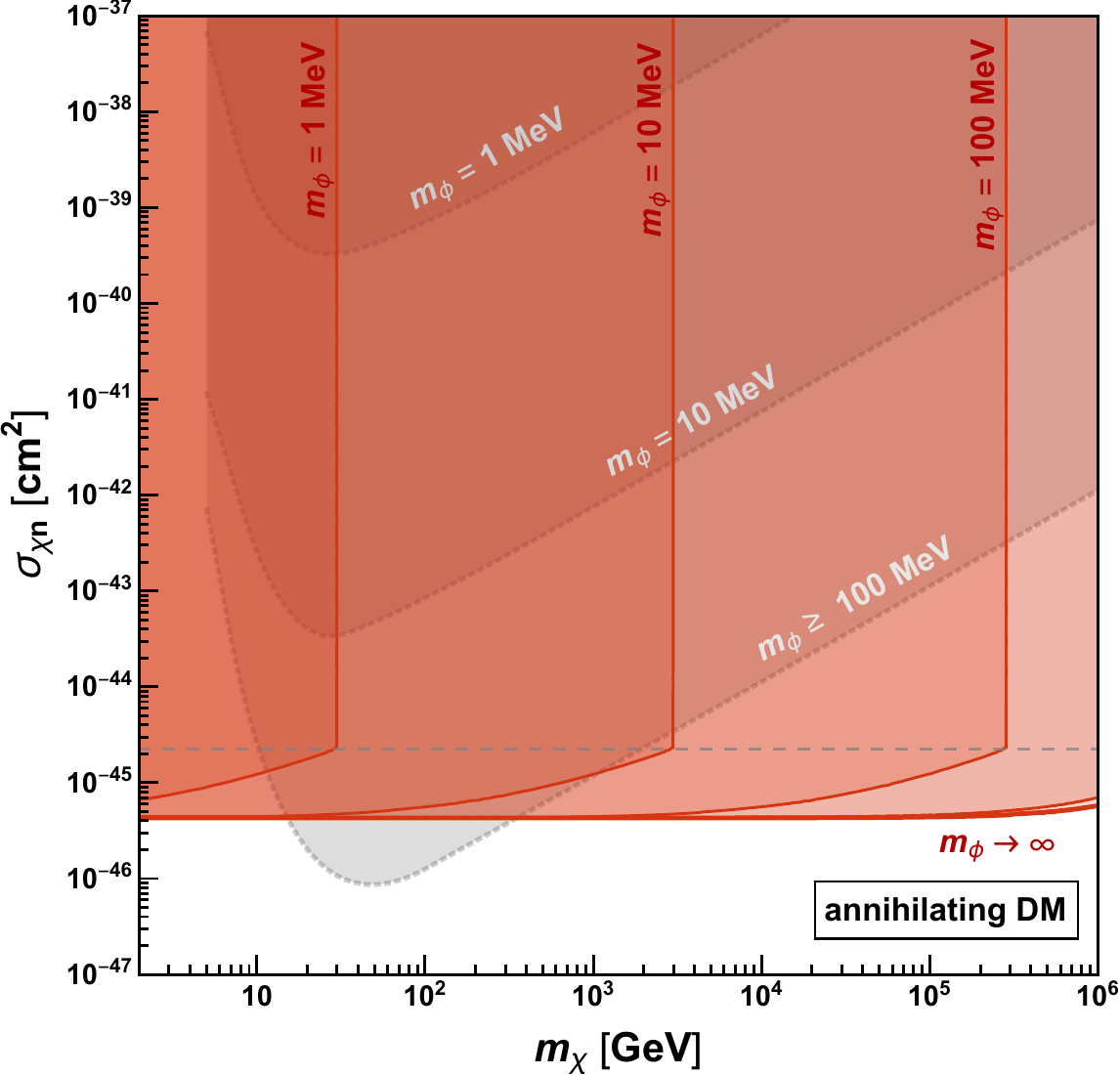}
\end{center}
	\caption{Projected upper limits on DM-nucleon scattering cross section for annihilating DM, obtainable from dark heating of a neutron star with surface temperatures of 1950\,K, shown for different mediator masses. Red shaded regions above the solid red lines are excluded; lines corresponding to mediator masses of  $m_{\phi}= 1 $ MeV, $m_{\phi}= 10 $ MeV, $m_{\phi}= 100 $ MeV, respectively, from left to right. For  $m_{\phi}= 1 $ GeV, the constraint is close to the case with $m_{\phi}\to \infty$. Related spin-independent exclusion limits from the underground detector {\sc PandaX-II}~\cite{Ren:2018gyx} are shown in the grey shaded regions above dotted grey lines. For mediators heavier than 100 MeV, exclusion limits from direct detection experiments are close to those obtained assuming an infinitely massive mediator. The dashed grey horizontal line corresponds to the geometrical saturation cross section above which any cross section is essentially equivalent to saturation cross section and therefore ruled out with same confidence. The local DM density around the neutron star is taken as 0.4 GeV/$\rm cm^3$ and all the other neutron star parameters are described in the text. We restrict our study to $m_{\chi}<10^{6}$\,GeV to ensure multiple scattering is not relevant. As evident, the dark heating constraints weaken significantly with lighter mediators, being washed out for $m_{\phi} \leq 0.25 $ MeV. However, the astrophysical limits and the terrestrial limits weaken in a complementary fashion.}
	\label{fig:ann}
\end{figure}

\begin{figure}
	\centering
	\includegraphics[scale=0.5]{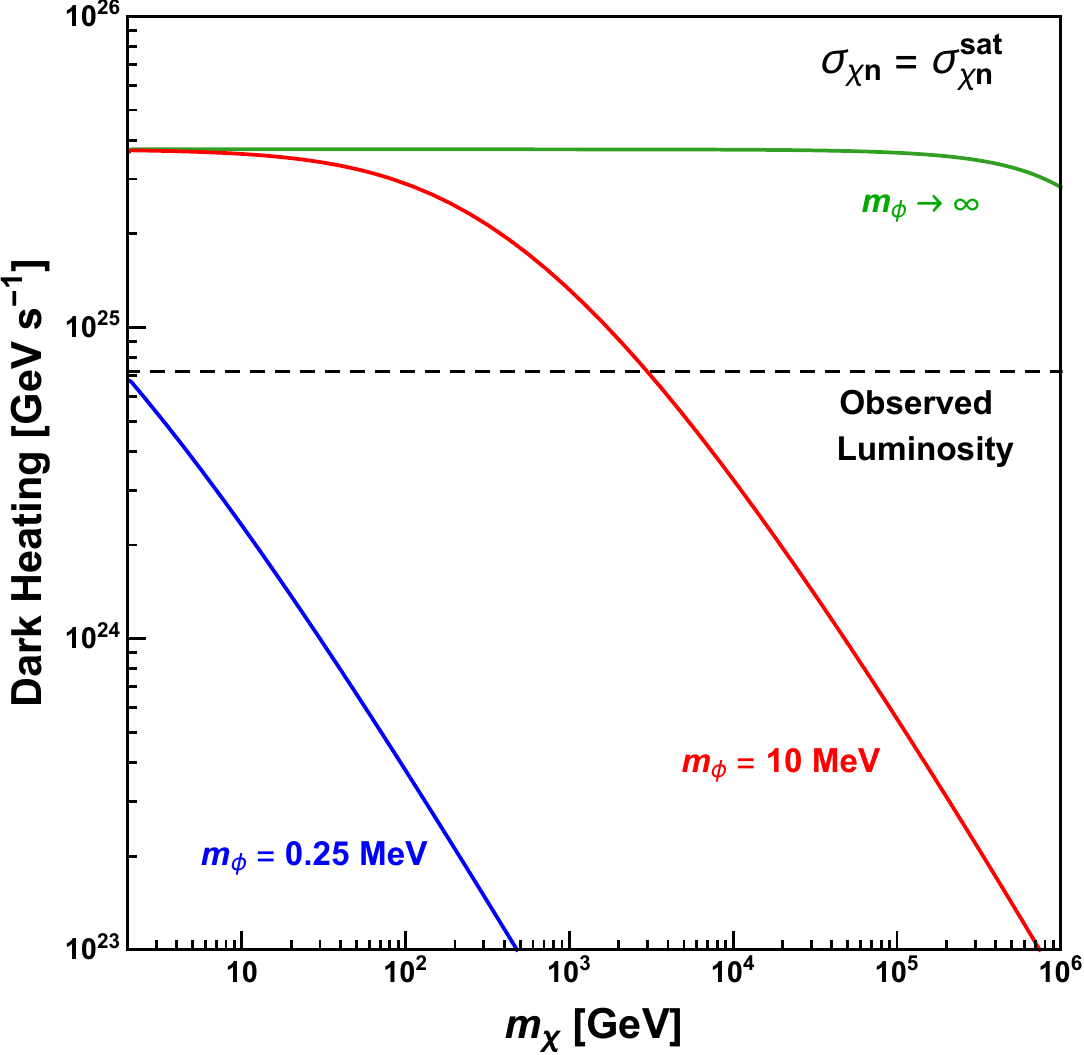}
	\caption{Dark heating from annihilation of captured DM particles \emph{\&} kinetic heating is shown with dark matter mass for a neutron star, $\sigma^{\rm sat}_{\chi \rm n}$ denotes the geometrical saturation value of DM-nucleon scattering cross section. Dashed black line corresponds to the luminosity of the  neutron star with surface temperature of 1950\,K. Green, red, and blue lines correspond to mediator masses of $m_{\phi} \to \infty $, $m_{\phi}= 10 $ MeV, and $m_{\phi}= 0.25 $ MeV respectively. The local DM density around the neutron star is taken as 0.4 GeV/$\rm cm^3$ and all the other neutron star parameters are described in the text. We restrict our study to $m_{\chi}<10^{6}$\,GeV to ensure multiple scattering is not relevant. For mediators lighter than 0.25 MeV, dark heating can not exceed the observed luminosity. As a consequence, constraints on DM-nucleon scattering cross section from dark heating are completely washed out for mediators lighter than 0.25 MeV.}
	\label{fig:woheating}	
\end{figure}

For annihilating DM, the captured dark matter particles after being {thermalized} can annihilate among themselves, which in turn thermalize and heat up the neutron star. For such heating to be efficient, the annihilation products must have mean free paths smaller than the size of the star. All Standard Model (SM) final states satisfy this criterion. Further, within the parameter space of our interest, the thermalization timescale of captured DM particles, which depends solely on the DM-nucleon scattering cross section~\cite{McDermott:2011jp,Bell:2013xk}, is considerably less than typical age of the neutron star. The additional \textit{dark luminosity} $L_{\rm ann}$ is simply given by the total mass capture rate inside the neutron star, provided that the {equilibriation timescale} is less than the age of the neutron star. We have checked that within the parameter space shown in this work, {equilibriation time}~\cite{Bramante:2017xlb}  is extremely small compared to the typical neutron star lifetime and, as a consequence, dark luminosity simplifies to
\begin{equation}
L_{\rm ann} = m_{\chi}C_a N_{\chi}(t_{\rm age})^2 = m_{\chi}C\,.
\end{equation}

In addition to the dark luminosity, DM-nucleon scattering can also kinematically heat up the neutron star. DM particles acquire immense kinetic energies while falling into the steep gravitational potential of the neutron star, can transfer the kinetic energy to the nucleons during collisions, and as a consequence can heat up the neutron star. This is known as \textit{dark kinetic heating}, $L_{\rm kin}$~\cite{Baryakhtar:2017dbj,Raj:2017wrv}. For the neutron star parameters used in this work, the dark kinetic heating is given by  $L_{\rm kin} \sim 0.3 L_{\rm ann}$.
Therefore, a conservative upper limit on DM-nucleon interaction strength, as shown in Fig.\,\ref{fig:ann}, can simply be obtained by requiring that the dark heating not exceed  the observed luminosity 
\begin{equation}
L_{\rm ann} + L_{\rm kin} \, \leq L_{\rm obs}\,.
\end{equation}

In order to put stringent constraints on DM-nucleon interaction strength, observations of neutron stars with low enough surface temperatures or observations of neutron stars in DM rich environments are required. This is because the low surface temperature  $T_{\rm obs}$ of neutron star significantly reduces the observed luminosity as $L_{\rm obs} \sim  T^4_{\rm obs}$, and the large DM density in DM rich environments increases the dark heating that is proportional to the local DM density.  Right now, even with the coldest observed neutron star, PSR J2144–3933~\cite{Guillot:2019ugf}, dark heating can not competitively probe interesting DM-nucleon scattering cross sections. But, as discussed in~\cite{Baryakhtar:2017dbj}, radiation from a neutron star of surface temperature $\sim$ 1750\,K near the Earth can be detected by the upcoming telescopes, JWST~\cite{Gardner:2006ky}, TMT~\cite{Crampton:2008gx}, and E-ELT~\cite{Maiolino:2013bsa}. Possible detection of such neutron stars with imminent telescope technology, or alternatively detections of neutron stars with surface temperatures $\sim \mathcal{O} (10^4)$\,K in DM rich environments, typically project extremely stringent constraints on DM-nucleon scattering cross section~\cite{Bramante:2017xlb,Baryakhtar:2017dbj}. However, such strong upper limits entirely rely on the assumption of uniform energy loss distribution, and weaken significantly with lighter mediators as shown in Fig.\,\ref{fig:ann}.  From Fig.\,\ref{fig:ann}, it is also evident that, in the contact interaction approximation, constraints from dark heating are significantly stronger than terrestrial direct detection experiments for heavier DM mass. The reason behind this is also simple; constraints obtained from dark heating is essentially mass independent in the contact interaction approximation, whereas, constraints from direct detection experiments weaken proportionally to the DM mass for heavier DM. However, for interactions mediated via light mediators, the capture probability, and thereby the dark heating, decreases with lighter mediator.

The upper limits on DM-nucleon interaction strength,  as shown in Fig.\,\ref{fig:ann}, can be understood qualitatively. The flux of incoming  DM particles scales as $1/m_{\chi}$, the capture rate of DM particles is inversely proportional to the DM mass. As a consequence,  dark heating becomes independent of DM mass being essentially the total mass capture rate, explaining the $m_{\chi}$ independence of the upper limits. For higher DM mass, due to the kinematic suppression of a light mediator, the capture rate begins to scale as $\sim 1/m^2_{\chi}$, and as a consequence, dark heating scales as $\sim 1/m_{\chi}$, explaining the departure from $m_{\chi}$ independence. Eventually, once the DM-nucleon scattering cross section reaches the geometrical saturation value  the constraints become independent of $\sigma_{\chi \textrm{n}}$, explaining the sharp vertical cutoff. For a general energy loss distribution, constraints weaken with lighter mediators simply because the cutoff mass decreases as $\sim m^2_{\phi}$, which is evident from the generalized capture probability in Eq.\,(\ref{capture}).  

The astrophysical  exclusion limits weaken for mediators lighter than 1 GeV as shown in Fig.\,\ref{fig:ann}, in contrast to those
from underground detector that weaken for mediators lighter than 100 MeV. This has to do with the typical relative velocity in each case. For underground detectors, the average velocity of the DM particles is $\mathcal{O} (10^{-3})$ and the typical momentum transfer is in the range of 1-100 MeV for DM mass of 1\,GeV- 1\,PeV. As a consequence, the contact interaction approximation holds for mediators heavier than 100 MeV. For astrophysical constraints, the velocity of the DM particles is significantly enhanced while falling into the steep gravitational potential of the neutron star. Thus, the typical momentum transfer is $\sim 1$ GeV for DM mass above 1 GeV, thereby, invalidating the contact interaction approximation for much higher mediator mass compared to the underground direct detection experiments.

Fig.\,\ref{fig:woheating} illustrates why capture constraints on DM-nucleon scattering cross section of annihilating DM are significantly weakened for light mediators. For mediators lighter than 0.25 MeV, dark heating can not exceed the observed luminosity as shown in Fig.\,\ref{fig:woheating}, resulting a complete washout of the constraints on DM-nucleon scattering cross section. Observations of neutron stars with surface temperature of $\mathcal{O} (10^4)$\,K in DM rich environments also project similar constraints on DM-nucleon scattering cross section.

\subsection{Accretion of non-annihilating DM}

For asymmetric DM, the number of dark matter particles captured inside
the stellar object follows
\begin{equation}
\frac{dN_{\chi}}{dt} = C\,.
\end{equation}
Therefore, the number of  DM particles accumulating inside the neutron star increases linearly with time.  If the total number of captured DM particles inside the neutron star becomes self-gravitating, collapse ensues and if the zero point energy is not sufficient  to prevent the gravitational collapse, a black hole forms inside the neutron star. So, the \textit{collapse criterion} is simply given by~\cite{McDermott:2011jp,Garani:2018kkd} 
\begin{equation}\label{blackhole}
N_{\chi}(t_{\rm age}) \geq {\rm Max}\left [N^{\rm self}_{\chi}, N^{\rm cha}_{\chi} \right]\,.
\end{equation}

\begin{figure}
\centering	
\includegraphics[angle=0.0,width=0.79\textwidth]{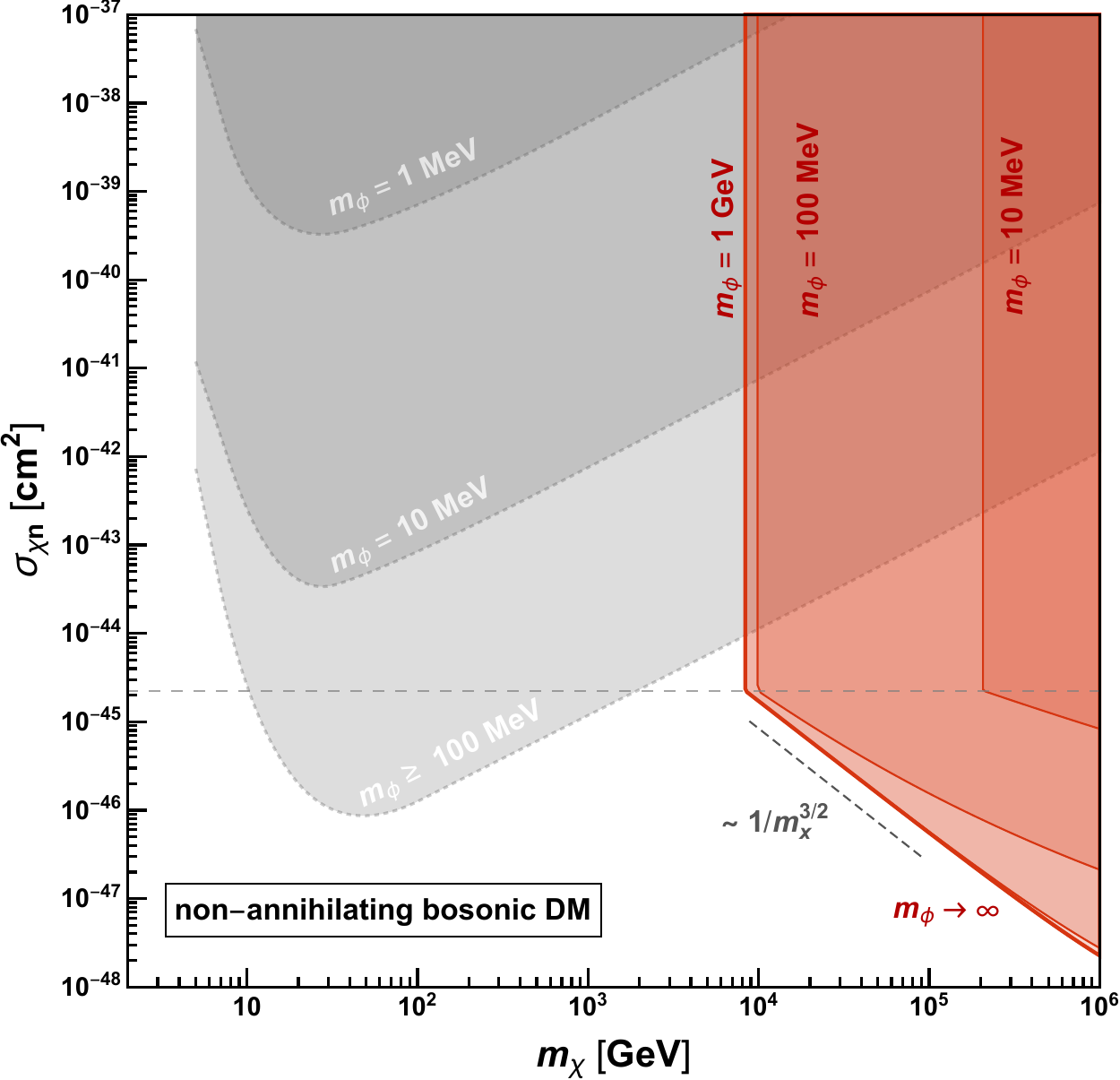}
	\caption{Upper limits on DM-nucleon scattering cross section for a bosonic non-annihilating DM, obtained from existence (i.e., non-collapse) of an old neutron star, PSR J0437-4715, shown for different mediator masses. Regions above the solid red lines are excluded. The lines correspond to mediator masses of  $m_{\phi}= 1 $ GeV, $m_{\phi}= 100 $ MeV, and $m_{\phi}= 10$\,MeV, respectively, from left to right. For  $m_{\phi}= 1 $ GeV, the constraint is close to the case with $m_{\phi}\to \infty$. Related spin-independent exclusion limits from the underground detector {\sc PandaX-II}~\cite{Ren:2018gyx} are shown in the grey shaded regions above dotted grey lines. For mediators heavier than 100 MeV, exclusion limits from direct detection experiments remain unaltered.  The dashed grey horizontal line corresponds to the geometrical saturation cross section above which any cross section is essentially equivalent to saturation cross section and therefore ruled out alike. The local DM density around the neutron star is taken as 0.4 GeV/$\rm cm^3$ as it is 139 $\pm$ 3 pc from the solar system and all the other neutron star parameters are described in the text. We restrict our study to $m_{\chi}<10^{6}$\,GeV to ensure multiple scattering is not relevant, but, in any case, the bounds are cut off beyond $m_{\chi}\gtrsim3\times10^{7}$\,GeV. Constraints on DM-nucleon scattering cross section obtained from gravitational collapse weaken significantly for lighter mediators, being washed out for $m_{\phi}\leq 5$\,MeV. Terrestrial limits also weaken, but in a complementary fashion. Astrophysical constraints extend only upto $30$\,PeV due to efficient Hawking evaporation.}
	\label{fig:collapse}	
\end{figure}

\begin{figure}
	\centering
	\includegraphics[scale=0.53]{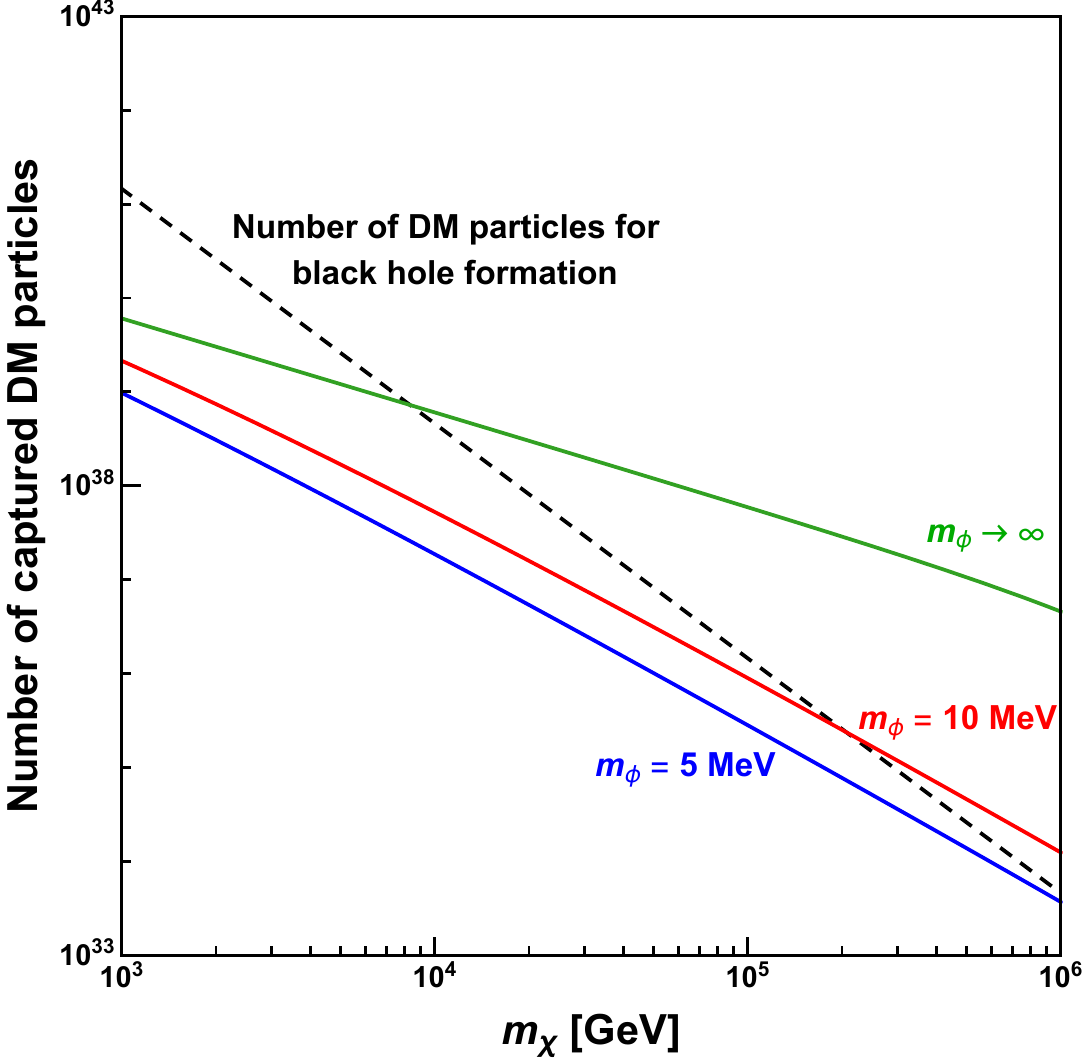}
	\caption{Total number of captured DM particles is shown with dark matter mass for a relatively old neutron star, PSR J0437-4715. $\sigma^{\rm sat}_{\chi \rm n}$ denotes the geometrical saturation value of DM-nucleon scattering cross section.  Dashed black line corresponds to the number of DM particles required for black hole formation which is simply given by $\textrm{Max}\left [N^{\rm self}_{\chi}, N^{\rm cha}_{\chi} \right]$. Green, red, and blue lines correspond to mediator masses of \mbox{$m_{\phi} \to \infty $}, $m_{\phi}= 10 $ MeV, and $m_{\phi}= 5 $ MeV respectively. The local DM density around the neutron star is taken as 0.4 GeV/$\rm cm^3$ and all the other neutron star parameters are described in the text. We restrict our study to $m_{\chi}<10^{6}$\,GeV to ensure multiple scattering is not relevant, and, in any case, the bounds are cut off beyond $m_{\chi}\gtrsim3\times10^{7}$\,GeV. For mediators lighter than 5 MeV, total number of captured DM particles cannot exceed the number of DM particles required for black hole formation. As a consequence, constraints on DM-nucleon scattering cross section from gravitational collapse completely wash out for mediators lighter than 5 MeV.}
	\label{fig:wocollapse}	
\end{figure}

The black hole formed inside the neutron star, due to the gradual accretion of DM particles followed by a subsequent collapse, can accrete mass from the host neutron star and can even completely destroy the host. Therefore, the upper limits on  DM-nucleon interaction strength from gravitational collapse, arise simply from the existence of the neutron star and are shown in Fig.\,\ref{fig:collapse}. We have used the observation of a relatively old neutron star, PSR J0437-4715, in this analysis. The core temperature and age of PSR J0437-4715 is discussed in~\cite{McDermott:2011jp} and are given by $2.1 \times 10^6\,$ K and $6.69 \, \times 10^9$ years respectively. PSR J0437-4715 is located at a distance of about $139 \pm 3$ pc from the solar system and therefore, the local DM density around the neutron star is assumed as 0.4 $\rm GeV/cm^3$.  

The collapse condition in Eq.\,(\ref{blackhole})  essentially implies that the density of DM particles within the thermalized sphere has to be larger than the baryonic density within that volume. In order to satisfy this criterion, DM particles have to obey
\begin{equation}
 \rho_{\chi} = \frac{m_{\chi}N^{\rm self}_{\chi}}{ \left(\frac{4}{3} \pi r^3_{\rm th}\right) } \, \geq \, \rho_{\rm NS}\,. 
\end{equation}
However, the Chandrasekhar limit, $N^{\rm cha}_{\chi}$, depends on the  spin-statistics of DM particles. For bosonic DM, the zero point energy is supported by the uncertainty principle, whereas, for fermionic DM, it is supported by the Pauli exclusion principle. See~\cite{McDermott:2011jp} for a detailed estimation of Chandrasekhar limit for both bosonic and fermionic DM. In the absence of any repulsive self-interactions among the dark matter particles, the Chandrasekhar limit for bosonic dark matter, $\sim 1.5 \times 10^{34} \left(100 \,  \textrm {GeV}/m_{\chi}\right)^2$, is much less than that for fermionic dark matter, $\sim 1.8 \times 10^{51} \left(100\, \textrm {GeV}/m_{\chi}\right)^3$. Therefore, it is obvious that in the absence of any repulsive DM self-interactions, bosonic DM experiences gravitational collapse much more easily than fermionic DM. As a consequence, the corresponding constraints on DM-nucleon interaction strength from gravitational collapse are much more stringent for bosonic DM compared to that for fermionic DM. In this analysis, we will only consider collapse of asymmetric bosonic dark matter. For asymmetric bosonic DM, we found that within our parameter space of interest, black hole formation criterion is essentially driven by the self gravitation condition, i.e.,  $N^{\rm self}_{\chi} \ge N^{\rm cha}_{\chi}$, consistent with~\cite{McDermott:2011jp}. 

The upper limits  shown in Fig.\,\ref{fig:collapse} can also be qualitatively understood.  The total number of captured DM particles inside the neutron star is proportional $1/m_{\chi}$, and in the collapse criterion, $N_{\chi}(t_\textrm{age})\geq\textrm{Max}[N_{\chi}^\textrm{self},N_{\chi}^\textrm{cha}]$, the right hand side is essentially dominated by self-gravitation, $N^{\rm self}_{\chi}$, which has a $1/m^{5/2}_{\chi}$ dependence. This explains the $1/m^{3/2}_{\chi}$ dependence of the upper limits shown in Fig.\,\ref{fig:collapse}. When the scattering cross section reaches its geometrical saturation value, the constraints become independent of $\sigma_{\chi \textrm{n}}$, which explains the sharp vertical cutoff. For lighter mediators, the capture probability and the total number of captured DM particles, decreases, explaining the weakening of the constraints compared to the upper limit obtained from a contact interaction approximation.
From Fig.\,\ref{fig:collapse}, it is clear that even in the absence of any repulsive self-interactions among the DM particles the astrophysical upper limits on DM-nucleon scattering cross section weaken appreciably with lighter mediators. For mediators lighter than 5 MeV, the total  number of captured DM particles is not sufficient to form a black hole, as shown in Fig.\,\ref{fig:wocollapse}. Hence, such astrophysical constraints are completely washed out. 

Note that if the newly formed black hole evaporates via Hawking radiation faster than it accretes mass, the host neutron star can still survive. As a consequence, the constraints obtained from gravitational collapse can be alleviated~\cite{McDermott:2011jp,Kouvaris:2010jy,Bell:2013xk,Garani:2018kkd}. The mass accretion rate of the newly formed black hole  is given by
 \begin{equation}
 \frac{dM_{\rm acc}}{dt} = \frac{4\pi \lambda \rho_{\rm NS} \,G^2_{\rm N}\, M^2_{\rm BH}}{c^3_{\rm s}}\,,
 \end{equation}
 where $\lambda = 0.25$, $c_{\rm s} = 0.17c$ is the speed of sound inside the neutron star, $G_{\rm N}$ denotes gravitational constant, and $M_{\rm BH}$ is the mass of the newly formed black hole~\cite{Bell:2013xk}. 
 
In all previous treatments, the relevant Hawking evaporation rate is calculated by treating the newly formed black hole as a blackbody. This is, of course, an approximation. The spectrum of the emitted particles due to Hawking radiation is not exactly a blackbody spectrum, but rather follows~\cite{Hawking:1974rv,Hawking:1974sw,MacGibbon:1990zk,MacGibbon:1991tj,Dasgupta:2019cae}
 \begin{equation}
 \frac{d^2N}{dEdt} = \frac{1}{2\pi\hbar} \frac{\Gamma_{\rm s}(E,M_{\rm{BH}})}{{\exp}\left[{E}/{T_{\rm{BH}}} \right]-(-1)^{2{\rm s}}} \,,
 \label{eq:Differential energy distribution}
 \end{equation}  
where $s$ is the spin of emitted particle, $E$ is the energy of the emitted particle and $T_{\rm BH}$ denotes the temperature of the black hole. $\Gamma_{\rm s}(E,M_{\rm{BH}})$ denotes the dimensionless absorption probability, usually called the grey-body factor~\cite{MacGibbon:1990zk}. Therefore, the mass loss rate via Hawking evaporation, obtained by summing over all emitted species, is given by~\cite{MacGibbon:1991tj,Carr:2020gox}
 \begin{equation}
 \frac{dM_{10}}{dt} = -5.34 \times 10^{-5} \,\frac{f(M)}{M^2_{10}}\,,
 \end{equation}
 where $M_{10} = M_{\rm BH}/ \left(10^{10}\,\rm g\right)$ and $f(M)$ measures the number of emitted particle species. Summing over contributions from all SM particles, $f(M)$ corresponds to 15.35~\cite{Carr:2020gox}. As a consequence, the correct Hawking evaporation rate of the newly formed black hole simplifies to
 \begin{equation}
 \frac{dM_{\rm evap}}{dt} = - \frac{\hbar\,c^4}{74\pi G^2_{\rm N}M^2_{\rm BH}}\,.
 \end{equation}
 
Constraints on DM-nucleon cross section get alleviated when the Hawking evaporation rate exceeds the mass accretion rate, i.e., if $dM_{\rm acc}/dt \geq dM_{\rm evap}/dt$. For a relatively cold neutron star, PSR J0437-4715, with core temperature of $2.1 \times 10^6$\,K,  we estimate that such astrophysical constraints on DM-nucleon interaction strength obtained from gravitational collapse vanish at $m_{\chi} \geq 3 \times 10^7$ GeV for bosonic DM and $m_{\chi} \geq 8 \times 10^9$ GeV for fermionic DM.  Therefore, the constraints shown in Fig.\,\ref{fig:collapse} can only extend up to $3 \times 10^7$ GeV. 
 
\subsection{Effect of DM self-interactions on bounds on DM-nucleon interactions}
\label{sec:selfint}

For annihilating DM, constraints on DM-nucleon scattering cross section are simply obtained by comparing the dark heating with the observed luminosity. The estimation of dark heating solely depends on the total capture rate. In principle, self-interactions among the DM particles can lead to an additional contribution to the total capture rate. However, due to the extremely large baryonic density in neutron stars, self-interactions have an insignificant impact on the total capture rate. Therefore, the upper limits shown in Fig.\,\ref{fig:ann} remain unaltered even with the inclusion of maximally allowed repulsive self-interactions among the DM particles.

\begin{figure}
	\centering
	\includegraphics[angle=0.0,width=0.45\textwidth]{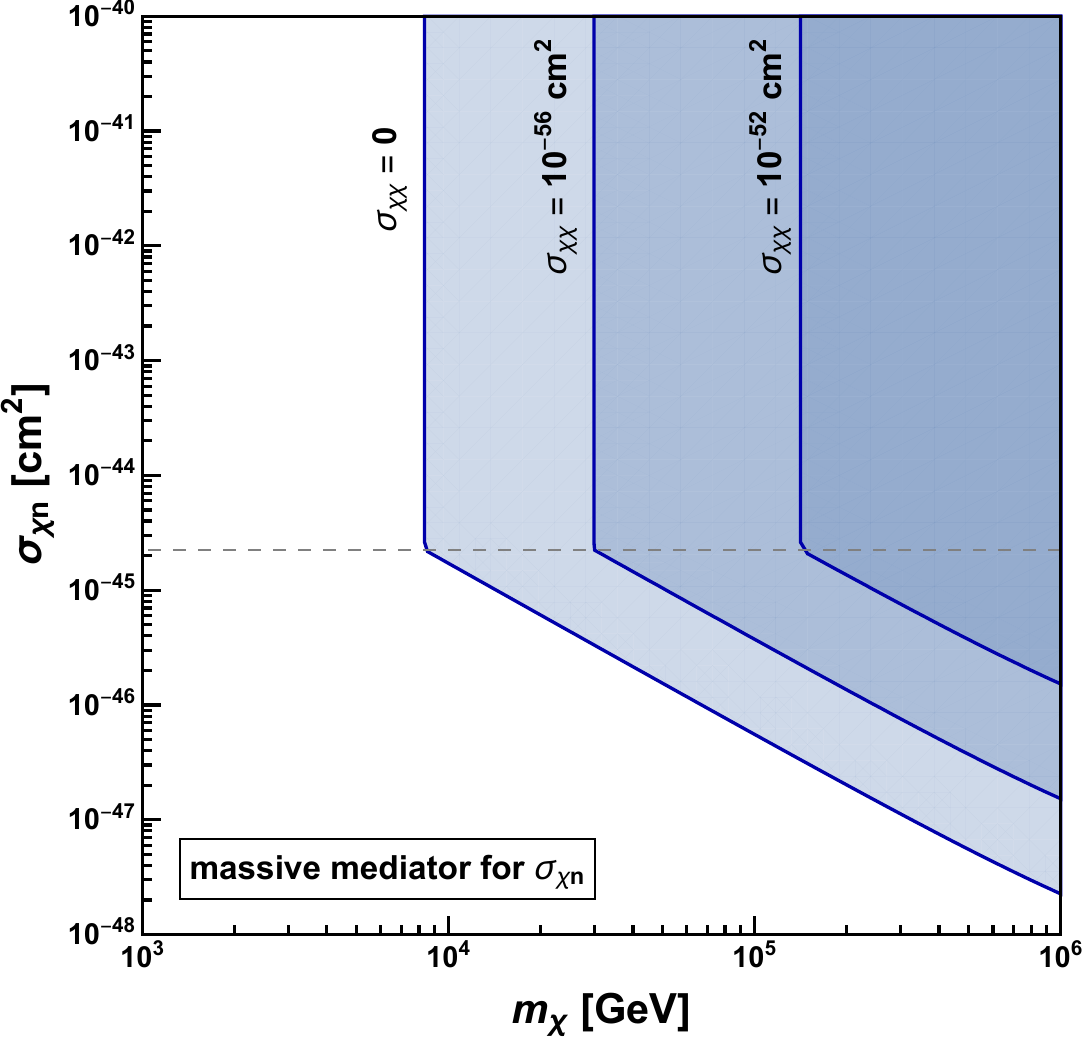}\quad\quad
	\includegraphics[angle=0.0,width=0.45\textwidth]{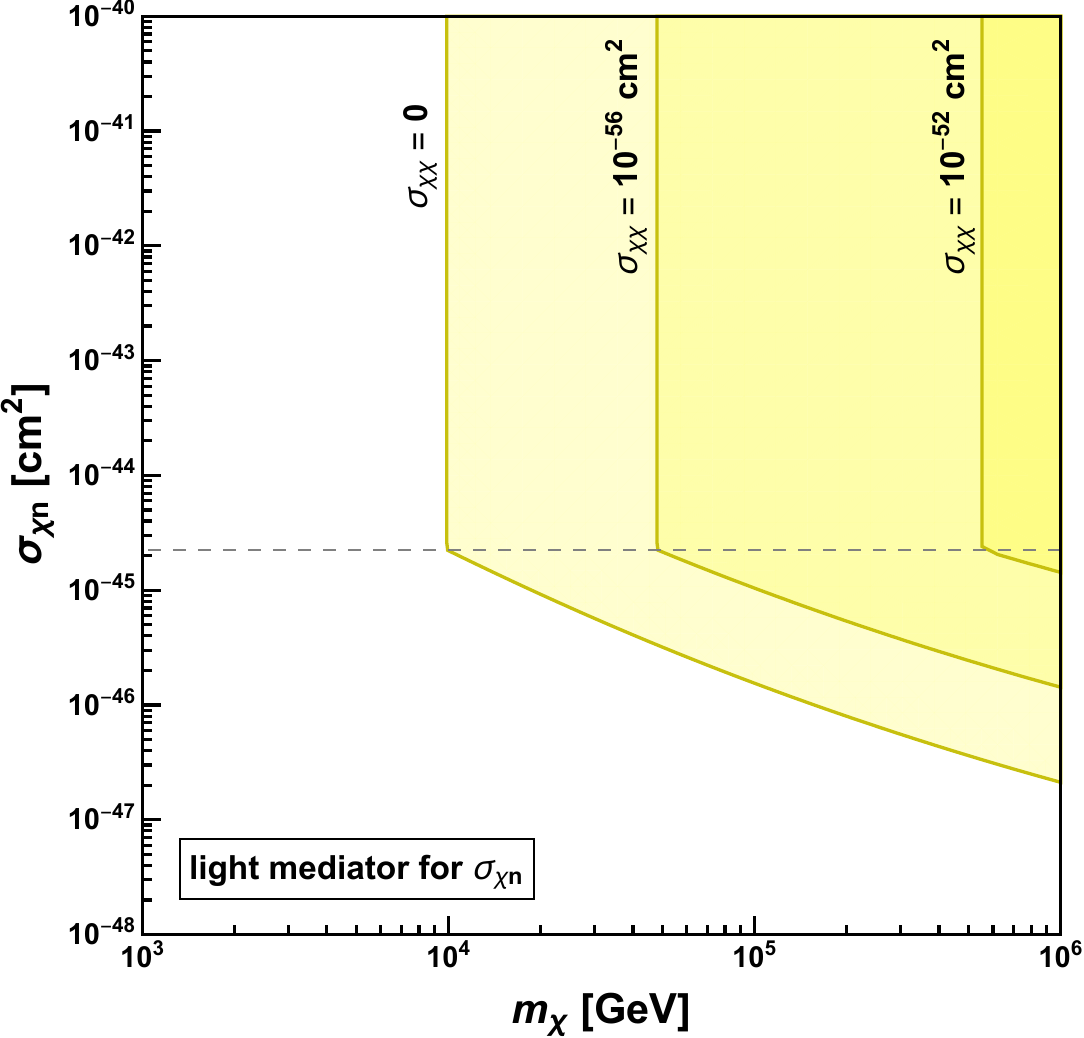}
	\caption{Upper limits on DM-nucleon scattering cross section of non-annihilating self-repelling bosonic DM, obtained from non-observation of gravitational collapse of PSR J0437-4715. In the left panel, a uniform energy loss  distribution is assumed, and in the right panel the DM-nucleon interaction is taken to be mediated via a Yukawa potential with a mediator mass of 100 MeV. In each panel, different lines correspond to  $\sigma_{\chi \chi}=0$, $\sigma_{\chi \chi} = 10^{-56}\, \rm cm^2$, and $\sigma_{\chi \chi} = 10^{-52}\, \rm cm^2$, respectively, from left to right. The values of the DM self-interaction cross section shown in the plots, are not ruled out by the Bullet Cluster observation~\cite{Randall:2007ph}. The dashed grey line corresponds to the geometrical saturation cross section, above which any cross section is essentially equivalent to saturation cross section and therefore ruled out at same confidence. We restrict our study to $m_{\chi}<10^{6}$\,GeV to ensure multiple scattering is not relevant. Constraints on DM-nucleon scattering cross section weaken, and eventually washout, with increase of the strength of repulsive self-interactions.}
	\label{fig:self}	
\end{figure}

For asymmetric DM, the inclusion of repulsive self-interactions among the DM particles has a  prominent impact on the corresponding upper limits on the DM-nucleon scattering cross section. In neutron stars, self-interactions among the DM particles do not enhance the capture rate but repulsive self-interactions among the dark matter particles significantly enhance the Chandrasekhar limit, thereby, increasing the number of DM particles required for black hole formation. For a repulsive self-interaction of contact type, the Chandrasekhar limit for bosonic DM modifies to~\cite{Bell:2013xk}
\begin{equation}\label{chand}
N^{\rm cha}_{\chi} = \frac{2}{\pi}\frac{M^2_{\rm pl}}{m^2_{\chi}} \left(1+\frac{\lambda}{32\pi} \frac{M^2_{\rm pl}}{m^2_{\chi}} \right)^{1/2}\,,
\end{equation}
where $\lambda$ is the dimensionless self-interaction coupling among the DM particles which is given by $\lambda = \left(64 \pi \sigma_{\chi \chi} \right)^{1/2} m_{\chi}$. Note that, the second term in the expression of $N^{\rm cha}_{\chi}$ always dominates over unity for any reasonable value of DM self scattering cross section. As a consequence, the corresponding constraints on DM-nucleon interaction strength further weaken with inclusion of repulsive DM self-interactions~\cite{Bell:2013xk,Bramante:2013hn}. We show the impact of repulsive DM self-interactions on the corresponding upper limits on DM-nucleon scattering cross section in Fig.\,\ref{fig:self}. From Fig.\,\ref{fig:self},  it is evident that with increase of the self-interaction strength, the Chandrasekhar limit significantly enhances and as a consequence, the black hole formation criterion becomes stricter. This explains the further weakening of the constraints. However, inclusion of attractive self-interactions among bosonic DM particles $(\lambda < 0)$ requires a different analysis involving bound state formation, which is beyond the scope of this work. For fermionic DM, the impact of attractive self-interactions among the DM particles on the  DM-nucleon scattering cross section have been extensively discussed in~\cite{Bramante:2013nma,Lin:2020zmm}.

\subsection{Constraining DM self-interactions from gravitational collapse}
\begin{figure}
\centering
	\includegraphics[angle=0.0,width=0.92\textwidth]{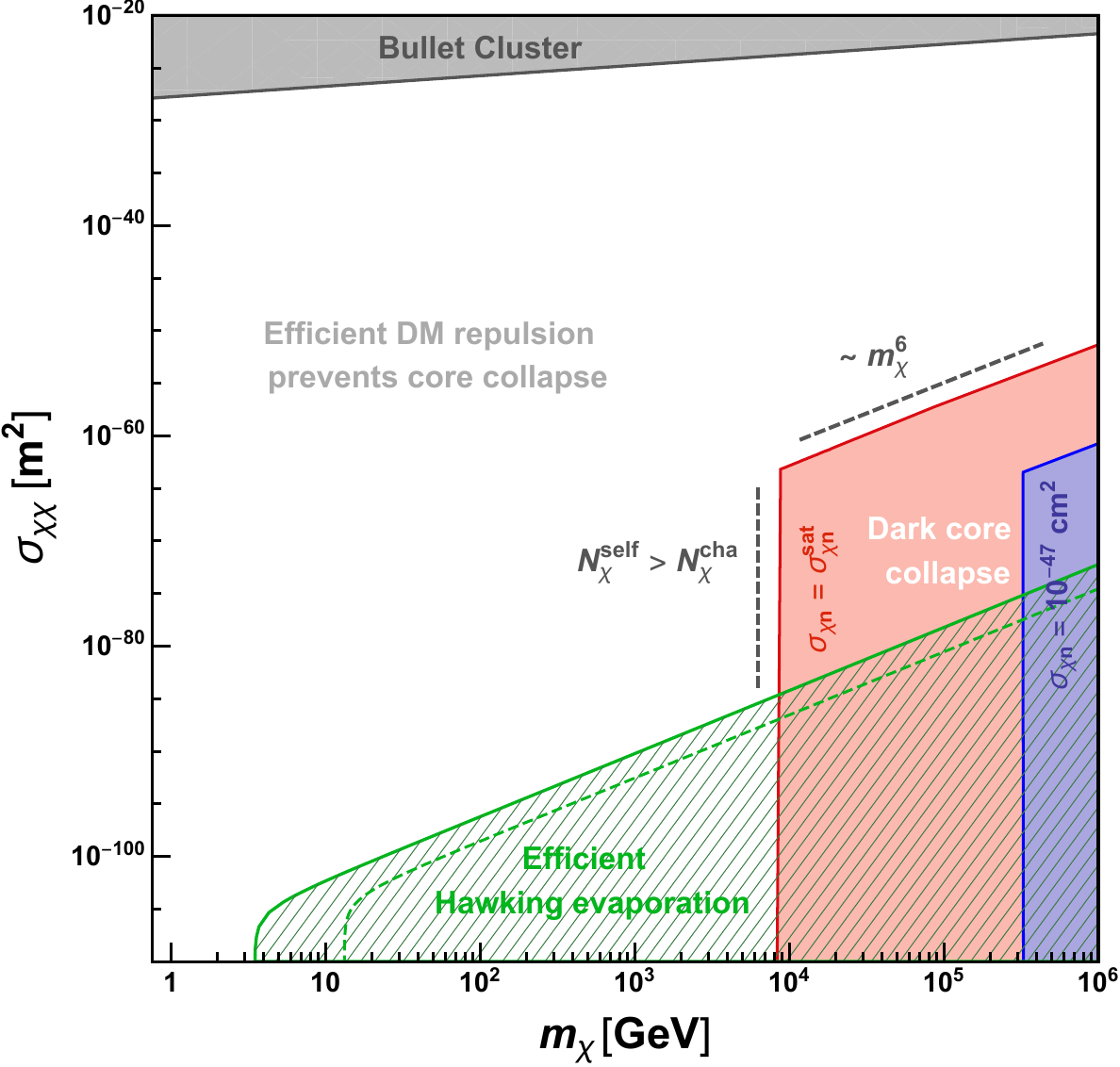}
	\caption{Constraints on DM self scattering cross section for  non-annihilating bosonic dark matter, obtained from a nearby old neutron star, PSR J0437-4715. The shaded regions denote parameter space where the DM core can collapse; red region for $\sigma_{\chi \textrm{n}} = \sigma^{\rm sat}_{\chi \textrm{n}} = 2\times 10^{-45} \, \rm cm^2$ and blue shaded region for  $\sigma_{\chi \textrm{n}} = 10^{-47} \rm \, cm^2$. Hatched region under the solid green line corresponds to the alleviation of the constraints due to rapid  Hawking evaporation of the newly formed black hole. Assuming a blackbody spectrum, instead of the correct Hawking spectrum, leads to the hatched region under the dashed green line. The exclusion limit from Bullet Cluster observation~\cite{Randall:2007ph} is shown in the grey shaded region at the top. The local DM density around the neutron star is taken as 0.4 GeV/$\rm cm^3$ and all the other neutron star parameters are described in the text. We restrict our study to $m_{\chi}<10^{6}$\,GeV to ensure multiple scattering is not relevant. Observation of a collapse event would rule out all parameter space that is white, unshaded and unhatched, without requiring knowledge of $\sigma_{\chi \textrm{n}}$ if accretion of non-dark matter is bounded otherwise. If $\sigma_{\chi \textrm{n}}$ is known, the existence of the star rules out the corresponding shaded but unhatched region.}
	\label{fig:dmself}
\end{figure}

Gravitational collapse of stars can also be used as an astrophysical probe of DM self-interaction strength. If our particle physics model includes self-interactions, limited only by observations such as that of the Bullet cluster, observation of a collapse event gives information on $\sigma_{\chi\chi}$.

As we discussed before, strong self-interactions make the black hole formation criterion stricter, while its impact on the total capture rate is negligible. Therefore, strong self-interactions among DM particles inhibits collapse. However, very feeble self-interactions among the dark matter particles do not alter the Chandrasekhar limit appreciably and collapse can occur. Thus observation of collapse rules out strong self-interactions. In Fig.\,\ref{fig:dmself}, the red or blue shaded regions correspond to parameter spaces where the DM core in a neutron star can collapse, for $\sigma_{\chi\textrm{n}}$ being $\sigma_{\chi\textrm{n}}^\textrm{sat}$ or $10^{-47}$\,cm$^2$, respectively. We show these regimes for a uniform energy loss distribution. Evaporation of the newly formed black hole faster than its mass accretion rate invalidates the above-stated astrophysical constraints on DM self-interaction strength.  This region is shown by the green hatched region in Fig.\,\ref{fig:dmself}. Assuming a blackbody spectrum, instead of the correct Hawking spectrum we have used, leads to the hatched region under the dashed green line, underestimating the efficacy of Hawking evaporation. 

Even if we do not know $\sigma_{\chi\textrm{n}}$, a collapse observation will conservatively exclude enormous swathes of parameter space in Fig.\,\ref{fig:dmself}. The entire white unhatched unshaded region is disfavored by the observed collapse of a star such as PSR J0437-4715, assuming that we can separately constrain the accretion of gas etc. If on the other hand, the value of $\sigma_{\chi\textrm{n}}$ is known, one can obtain bounds from non-observation of collapse. For illustration, if it is known that $\sigma_{\chi\textrm{n}}=10^{-47}$\,cm$^2$, existence of PSR J0437-4715 rules out the blue region that does not over lap with the hatched Hawking constraint.

The upper limits shown in Fig.\,\ref{fig:dmself} can also be understood qualitatively. The black hole formation criterion is $N_{\chi}(t_\textrm{age})\geq\textrm{Max}[N_{\chi}^\textrm{self},N_{\chi}^\textrm{cha}]$. The left hand side scales as $1/m_{\chi}$ and has no dependence on $\sigma_{\chi\chi}$, whereas for heavier DM the right hand side is dominated by the Chandrasekhar limit, which scales as $\sim \sigma^{1/4}_{\chi \chi}/ m^{5/2}_{\chi}$, so that the constraint scales as $\sigma_{\chi\chi}\sim m^6_{\chi}$. For lighter DM, $N_{\chi}^\textrm{self}$ dominates over $N_{\chi}^\textrm{cha}$ on the right hand side, and both $N_{\chi}^\textrm{self}$ and $N_{\chi}(t_\textrm{age})$ are independent of $\sigma_{\chi\chi}$, explaining the $\sigma_{\chi \chi}$ independence of the upper limit, i.e., the sharp vertical cutoff in Fig.\,\ref{fig:dmself}. These two regimes meet almost discontinuously at the corner of the red or blue shaded regions, where $N_{\chi}^\textrm{self}=N_{\chi}^\textrm{cha}$, identifying a minimum DM mass below which collapse cannot occur for a chosen $\sigma_{\chi\textrm{n}}$ and other fixed parameters. Note that, since the total number of captured DM particles is proportional to the DM-nucleon scattering cross section, lower values of DM-nucleon scattering cross section reduce the total number of captured DM particles, and as a consequence, enfeeble the upper limits on DM self-interaction strength as shown in Fig.\,\ref{fig:dmself}. 

\section{Summary \textit{\&} Outlook}
\label{sec:summ}
Dark matter capture in celestial objects is believed to be one of the most sensitive astrophysical probes of interactions between the dark and visible sector. However, all such previously obtained astrophysical constraints on DM-nucleon scattering cross section depend crucially on the assumption of a uniform energy loss distribution. This assumption of uniformity, tantamount to assuming an infinitely massive mediator, does not always hold true and the constraints must be updated for lighter mediators, self-interactions, and to better codify the effect of Hawking evaporation. Here we summarize our main results:

\begin{itemize}
	\item We pointed out that the assumption of uniformity in energy loss distribution only holds true for isotropic differential cross section, which occurs only for an infinitely massive mediator. For DM interactions via light mediators, the deviation from uniformity becomes prominent and is shown in Fig.\,\ref{fig:gen}.
	
	\item We generalized the  treatment of DM capture inside celestial objects for arbitrary mediator masses, in Eqs.\,(\ref{g1def} -- \ref{pp}), and updated the existing and projected astrophysical upper limits on DM-nucleon interaction strength.
		
	\item We found that for non-annihilating bosonic DM, the existing constraints obtained from an old nearby neutron star, PSR J0437-4715, and for annihilating DM, the projected constraints obtainable from a neutron star with surface temperature of 1950\,K, depend firmly on mediator masses, weaken appreciably for lighter mediators, and are not generally superior to terrestrial detectors as shown in Fig.\,\ref{fig:ann} and Fig.\,\ref{fig:collapse}, respectively. Such astrophysical constraints are completely washed out for mediators lighter than 5 MeV for asymmetric DM, and for mediators lighter than 0.25 MeV for annihilating DM as shown in Fig.\,\ref{fig:woheating} and Fig.\,\ref{fig:wocollapse}, respectively.
	
	\item We showed in Fig.\,\ref{fig:self} that the constraints on DM-nucleon interaction strength obtained from gravitational collapse further weaken with inclusion of repulsive self-interactions among the DM particles. As repulsive DM self-interactions have a significant impact on the black hole formation criterion, constraints can even be completely washed out for strong enough repulsive self-interactions among the DM particles. However, constraints obtained from dark heating remain unaltered with the inclusion of  DM self-interactions due to their insignificant contribution to the total capture rate.
	
	\item Gravitational collapse of captured DM particles  acts as an astrophysical probe of DM self scattering cross section. Self-interactions among the DM particles have an insignificant impact on the total capture rate but repulsive DM self-interactions have a significant impact on the black formation criterion. Strong repulsive DM self-interactions can prevent the black hole formation inside a stellar object. Black hole formation criterion can only be achieved with very feeble repulsive DM self-interaction strength. Thereby, from collapse of a stellar object the white region shown in Fig.\,\ref{fig:dmself} is ruled out. If $\sigma_{\chi\textrm{n}}$ is known, then existence of such stellar objects rule out shaded but unhatched regions therein.
	
	\item In all the previous treatments, the Hawking evaporation rate from the newly formed black hole was calculated in a blackbody radiation approximation. This is not completely correct. We correct the Hawking evaporation estimate, which leads to a stronger alleviation of the constraints on DM self-interaction strength, compared to what would be obtained using a blackbody approximation.
	
	\end{itemize}

Although we have explicitly applied our generalized treatment to update the astrophysical constraints on DM-nucleon interaction strength, it can easily be translated to update the existing constraints on leptophilic coupling of DM~\cite{Garani:2019fpa,Joglekar:2019vzy,Bell:2019pyc,Joglekar:2020liw}. Our generalized treatment can also be applied for various other NS models (e.g., BSK-20, BSK-21) to determine all the astrophysical constraints instead of assuming a uniform density~\cite{Garani:2018kkd}. For compact objects, due to their extremely large baryonic density, the contribution to capture rate from self-capture is insignificant. However, for non-compact celestial objects, self-capture plays a pivotal role in the estimate of the total capture rate~\cite{Zentner:2009is,Gaidau:2018yws}. Therefore, the generalized self-capture rate, which we have estimated in this work, will be relevant in the context of such non-compact stellar objects. We have not attempted to treat multiple scattering, light mediators, and self-capture within a single framework; it can be relevant for $m_{\chi}>10^{6}$\,GeV and remains to be addressed.

\section*{Acknowledgements}
We thank Triparno Bandyopadhyay, Raghuveer Garani, Subhajit Ghosh, Thomas Hambye, and Ranjan Laha for discussions and useful suggestions. The work of B.D. is supported by the Dept.\,\,of Atomic Energy (Govt.\,\,of India) research project 12-R\textit{\&}D-TFR-5.02-0200, the Dept.\,\,of Science and Technology (Govt.\,\,of India) through a Ramanujan Fellowship, and by the Max-Planck-Gesellschaft through a Max Planck Partner Group.   The work of A.G. is supported by the ``Probing Dark Matter through Neutrinos'' ULB-ARC grant.  

\bibliographystyle{JHEP}
\bibliography{ref}

\providecommand{\href}[2]{#2}\begingroup\raggedright\begin{thebibliography}{100}

\bibitem{Aghanim:2018eyx}
{\scshape Planck} collaboration, N.~Aghanim et~al., \emph{{Planck 2018 results.
  VI. Cosmological parameters}},
  \href{https://doi.org/10.1051/0004-6361/201833910}{\emph{Astron. Astrophys.}
  {\bfseries 641} (2020) A6}
  [\href{https://arxiv.org/abs/1807.06209}{{\ttfamily 1807.06209}}].

\bibitem{Jungman:1995df}
G.~Jungman, M.~Kamionkowski and K.~Griest, \emph{{Supersymmetric dark matter}},
  \href{https://doi.org/10.1016/0370-1573(95)00058-5}{\emph{Phys. Rept.}
  {\bfseries 267} (1996) 195}
  [\href{https://arxiv.org/abs/hep-ph/9506380}{{\ttfamily hep-ph/9506380}}].

\bibitem{Steigman:2012nb}
G.~Steigman, B.~Dasgupta and J.~F. Beacom, \emph{{Precise Relic WIMP Abundance
  and its Impact on Searches for Dark Matter Annihilation}},
  \href{https://doi.org/10.1103/PhysRevD.86.023506}{\emph{Phys. Rev.}
  {\bfseries D86} (2012) 023506}
  [\href{https://arxiv.org/abs/1204.3622}{{\ttfamily 1204.3622}}].

\bibitem{Aprile:2018dbl}
{\scshape XENON} collaboration, E.~Aprile et~al., \emph{{Dark Matter Search
  Results from a One Ton-Year Exposure of XENON1T}},
  \href{https://doi.org/10.1103/PhysRevLett.121.111302}{\emph{Phys. Rev. Lett.}
  {\bfseries 121} (2018) 111302}
  [\href{https://arxiv.org/abs/1805.12562}{{\ttfamily 1805.12562}}].

\bibitem{Akerib:2016vxi}
{\scshape LUX} collaboration, D.~S. Akerib et~al., \emph{{Results from a search
  for dark matter in the complete LUX exposure}},
  \href{https://doi.org/10.1103/PhysRevLett.118.021303}{\emph{Phys. Rev. Lett.}
  {\bfseries 118} (2017) 021303}
  [\href{https://arxiv.org/abs/1608.07648}{{\ttfamily 1608.07648}}].

\bibitem{Cui:2017nnn}
{\scshape PandaX-II} collaboration, X.~Cui et~al., \emph{{Dark Matter Results
  From 54-Ton-Day Exposure of PandaX-II Experiment}},
  \href{https://doi.org/10.1103/PhysRevLett.119.181302}{\emph{Phys. Rev. Lett.}
  {\bfseries 119} (2017) 181302}
  [\href{https://arxiv.org/abs/1708.06917}{{\ttfamily 1708.06917}}].

\bibitem{Agnes:2018oej}
{\scshape DarkSide} collaboration, P.~Agnes et~al., \emph{{Constraints on
  Sub-GeV Dark-Matter–Electron Scattering from the DarkSide-50 Experiment}},
  \href{https://doi.org/10.1103/PhysRevLett.121.111303}{\emph{Phys. Rev. Lett.}
  {\bfseries 121} (2018) 111303}
  [\href{https://arxiv.org/abs/1802.06998}{{\ttfamily 1802.06998}}].

\bibitem{Ren:2018gyx}
{\scshape PandaX-II} collaboration, X.~Ren et~al., \emph{{Constraining Dark
  Matter Models with a Light Mediator at the PandaX-II Experiment}},
  \href{https://doi.org/10.1103/PhysRevLett.121.021304}{\emph{Phys. Rev. Lett.}
  {\bfseries 121} (2018) 021304}
  [\href{https://arxiv.org/abs/1802.06912}{{\ttfamily 1802.06912}}].

\bibitem{Abdelhameed:2019hmk}
{\scshape CRESST} collaboration, A.~H. Abdelhameed et~al., \emph{{First results
  from the CRESST-III low-mass dark matter program}},
  \href{https://doi.org/10.1103/PhysRevD.100.102002}{\emph{Phys. Rev.}
  {\bfseries D100} (2019) 102002}
  [\href{https://arxiv.org/abs/1904.00498}{{\ttfamily 1904.00498}}].

\bibitem{Armengaud:2019kfj}
{\scshape EDELWEISS} collaboration, E.~Armengaud et~al., \emph{{Searching for
  low-mass dark matter particles with a massive Ge bolometer operated
  above-ground}}, \href{https://doi.org/10.1103/PhysRevD.99.082003}{\emph{Phys.
  Rev.} {\bfseries D99} (2019) 082003}
  [\href{https://arxiv.org/abs/1901.03588}{{\ttfamily 1901.03588}}].

\bibitem{Abramoff:2019dfb}
{\scshape SENSEI} collaboration, O.~Abramoff et~al., \emph{{SENSEI:
  Direct-Detection Constraints on Sub-GeV Dark Matter from a Shallow
  Underground Run Using a Prototype Skipper-CCD}},
  \href{https://doi.org/10.1103/PhysRevLett.122.161801}{\emph{Phys. Rev. Lett.}
  {\bfseries 122} (2019) 161801}
  [\href{https://arxiv.org/abs/1901.10478}{{\ttfamily 1901.10478}}].

\bibitem{Agnese:2018col}
{\scshape SuperCDMS} collaboration, R.~Agnese et~al., \emph{{First Dark Matter
  Constraints from a SuperCDMS Single-Charge Sensitive Detector}},
  \href{https://doi.org/10.1103/PhysRevLett.122.069901,
  10.1103/PhysRevLett.121.051301}{\emph{Phys. Rev. Lett.} {\bfseries 121}
  (2018) 051301} [\href{https://arxiv.org/abs/1804.10697}{{\ttfamily
  1804.10697}}].

\bibitem{Chivukula:1989cc}
R.~S. Chivukula, A.~G. Cohen, S.~Dimopoulos and T.~P. Walker, \emph{{Bounds on
  Halo Particle Interactions From Interstellar Calorimetry}},
  \href{https://doi.org/10.1103/PhysRevLett.65.957}{\emph{Phys. Rev. Lett.}
  {\bfseries 65} (1990) 957}.

\bibitem{Raffelt:1999tx}
G.~G. Raffelt, \emph{{Particle physics from stars}},
  \href{https://doi.org/10.1146/annurev.nucl.49.1.163}{\emph{Ann. Rev. Nucl.
  Part. Sci.} {\bfseries 49} (1999) 163}
  [\href{https://arxiv.org/abs/hep-ph/9903472}{{\ttfamily hep-ph/9903472}}].

\bibitem{Cyburt:2002uw}
R.~H. Cyburt, B.~D. Fields, V.~Pavlidou and B.~D. Wandelt, \emph{{Constraining
  strong baryon dark matter interactions with primordial nucleosynthesis and
  cosmic rays}}, \href{https://doi.org/10.1103/PhysRevD.65.123503}{\emph{Phys.
  Rev.} {\bfseries D65} (2002) 123503}
  [\href{https://arxiv.org/abs/astro-ph/0203240}{{\ttfamily
  astro-ph/0203240}}].

\bibitem{Fayet:2006sa}
P.~Fayet, D.~Hooper and G.~Sigl, \emph{{Constraints on light dark matter from
  core-collapse supernovae}},
  \href{https://doi.org/10.1103/PhysRevLett.96.211302}{\emph{Phys. Rev. Lett.}
  {\bfseries 96} (2006) 211302}
  [\href{https://arxiv.org/abs/hep-ph/0602169}{{\ttfamily hep-ph/0602169}}].

\bibitem{Mack:2007xj}
G.~D. Mack, J.~F. Beacom and G.~Bertone, \emph{{Towards Closing the Window on
  Strongly Interacting Dark Matter: Far-Reaching Constraints from Earth's Heat
  Flow}}, \href{https://doi.org/10.1103/PhysRevD.76.043523}{\emph{Phys. Rev.}
  {\bfseries D76} (2007) 043523}
  [\href{https://arxiv.org/abs/0705.4298}{{\ttfamily 0705.4298}}].

\bibitem{Dvorkin:2013cea}
C.~Dvorkin, K.~Blum and M.~Kamionkowski, \emph{{Constraining Dark Matter-Baryon
  Scattering with Linear Cosmology}},
  \href{https://doi.org/10.1103/PhysRevD.89.023519}{\emph{Phys. Rev.}
  {\bfseries D89} (2014) 023519}
  [\href{https://arxiv.org/abs/1311.2937}{{\ttfamily 1311.2937}}].

\bibitem{Ali-Haimoud:2015pwa}
Y.~Ali-Haïmoud, J.~Chluba and M.~Kamionkowski, \emph{{Constraints on Dark
  Matter Interactions with Standard Model Particles from Cosmic Microwave
  Background Spectral Distortions}},
  \href{https://doi.org/10.1103/PhysRevLett.115.071304}{\emph{Phys. Rev. Lett.}
  {\bfseries 115} (2015) 071304}
  [\href{https://arxiv.org/abs/1506.04745}{{\ttfamily 1506.04745}}].

\bibitem{Gluscevic:2017ywp}
V.~Gluscevic and K.~K. Boddy, \emph{{Constraints on Scattering of keV–TeV
  Dark Matter with Protons in the Early Universe}},
  \href{https://doi.org/10.1103/PhysRevLett.121.081301}{\emph{Phys. Rev. Lett.}
  {\bfseries 121} (2018) 081301}
  [\href{https://arxiv.org/abs/1712.07133}{{\ttfamily 1712.07133}}].

\bibitem{Raj:2017wrv}
N.~Raj, P.~Tanedo and H.-B. Yu, \emph{{Neutron stars at the dark matter direct
  detection frontier}},
  \href{https://doi.org/10.1103/PhysRevD.97.043006}{\emph{Phys. Rev.}
  {\bfseries D97} (2018) 043006}
  [\href{https://arxiv.org/abs/1707.09442}{{\ttfamily 1707.09442}}].

\bibitem{Slatyer:2018aqg}
T.~R. Slatyer and C.-L. Wu, \emph{{Early-Universe constraints on dark
  matter-baryon scattering and their implications for a global 21 cm signal}},
  \href{https://doi.org/10.1103/PhysRevD.98.023013}{\emph{Phys. Rev.}
  {\bfseries D98} (2018) 023013}
  [\href{https://arxiv.org/abs/1803.09734}{{\ttfamily 1803.09734}}].

\bibitem{Boddy:2018kfv}
K.~K. Boddy and V.~Gluscevic, \emph{{First Cosmological Constraint on the
  Effective Theory of Dark Matter-Proton Interactions}},
  \href{https://doi.org/10.1103/PhysRevD.98.083510}{\emph{Phys. Rev.}
  {\bfseries D98} (2018) 083510}
  [\href{https://arxiv.org/abs/1801.08609}{{\ttfamily 1801.08609}}].

\bibitem{Bhoonah:2018wmw}
A.~Bhoonah, J.~Bramante, F.~Elahi and S.~Schon, \emph{{Calorimetric Dark Matter
  Detection With Galactic Center Gas Clouds}},
  \href{https://doi.org/10.1103/PhysRevLett.121.131101}{\emph{Phys. Rev. Lett.}
  {\bfseries 121} (2018) 131101}
  [\href{https://arxiv.org/abs/1806.06857}{{\ttfamily 1806.06857}}].

\bibitem{Xu:2018efh}
W.~L. Xu, C.~Dvorkin and A.~Chael, \emph{{Probing sub-GeV Dark Matter-Baryon
  Scattering with Cosmological Observables}},
  \href{https://doi.org/10.1103/PhysRevD.97.103530}{\emph{Phys. Rev.}
  {\bfseries D97} (2018) 103530}
  [\href{https://arxiv.org/abs/1802.06788}{{\ttfamily 1802.06788}}].

\bibitem{Cappiello:2018hsu}
C.~V. Cappiello, K.~C.~Y. Ng and J.~F. Beacom, \emph{{Reverse Direct Detection:
  Cosmic Ray Scattering With Light Dark Matter}},
  \href{https://doi.org/10.1103/PhysRevD.99.063004}{\emph{Phys. Rev.}
  {\bfseries D99} (2019) 063004}
  [\href{https://arxiv.org/abs/1810.07705}{{\ttfamily 1810.07705}}].

\bibitem{Bringmann:2018cvk}
T.~Bringmann and M.~Pospelov, \emph{{Novel direct detection constraints on
  light dark matter}},
  \href{https://doi.org/10.1103/PhysRevLett.122.171801}{\emph{Phys. Rev. Lett.}
  {\bfseries 122} (2019) 171801}
  [\href{https://arxiv.org/abs/1810.10543}{{\ttfamily 1810.10543}}].

\bibitem{Krnjaic:2019dzc}
G.~Krnjaic and S.~D. McDermott, \emph{{Implications of BBN Bounds for Cosmic
  Ray Upscattered Dark Matter}},
  \href{https://doi.org/10.1103/PhysRevD.101.123022}{\emph{Phys. Rev. D}
  {\bfseries 101} (2020) 123022}
  [\href{https://arxiv.org/abs/1908.00007}{{\ttfamily 1908.00007}}].

\bibitem{Gluscevic:2019yal}
V.~Gluscevic et~al., \emph{{Cosmological Probes of Dark Matter Interactions:
  The Next Decade}},  \href{https://arxiv.org/abs/1903.05140}{{\ttfamily
  1903.05140}}.

\bibitem{Wadekar:2019xnf}
D.~Wadekar and G.~R. Farrar, \emph{{First direct astrophysical constraints on
  dark matter interactions with ordinary matter at very low velocities}},
  \href{https://arxiv.org/abs/1903.12190}{{\ttfamily 1903.12190}}.

\bibitem{Nadler:2019zrb}
E.~O. Nadler, V.~Gluscevic, K.~K. Boddy and R.~H. Wechsler, \emph{{Constraints
  on Dark Matter Microphysics from the Milky Way Satellite Population}},
  \href{https://doi.org/10.3847/2041-8213/ab1eb2}{\emph{Astrophys. J. Lett.}
  {\bfseries 878} (2019) 32}
  [\href{https://arxiv.org/abs/1904.10000}{{\ttfamily 1904.10000}}].

\bibitem{Press:1985ug}
W.~H. Press and D.~N. Spergel, \emph{{Capture by the sun of a galactic
  population of weakly interacting massive particles}},
  \href{https://doi.org/10.1086/163485}{\emph{Astrophys. J.} {\bfseries 296}
  (1985) 679}.

\bibitem{Gould:1987ju}
A.~Gould, \emph{{{WIMP} Distribution in and Evaporation From the Sun}},
  \href{https://doi.org/10.1086/165652}{\emph{Astrophys. J.} {\bfseries 321}
  (1987) 560}.

\bibitem{Gould:1987ir}
A.~Gould, \emph{{Resonant Enhancements in WIMP Capture by the Earth}},
  \href{https://doi.org/10.1086/165653}{\emph{Astrophys. J.} {\bfseries 321}
  (1987) 571}.

\bibitem{Bramante:2017xlb}
J.~Bramante, A.~Delgado and A.~Martin, \emph{{Multiscatter stellar capture of
  dark matter}}, \href{https://doi.org/10.1103/PhysRevD.96.063002}{\emph{Phys.
  Rev.} {\bfseries D96} (2017) 063002}
  [\href{https://arxiv.org/abs/1703.04043}{{\ttfamily 1703.04043}}].

\bibitem{Dasgupta:2019juq}
B.~Dasgupta, A.~Gupta and A.~Ray, \emph{{Dark matter capture in celestial
  objects: Improved treatment of multiple scattering and updated constraints
  from white dwarfs}},
  \href{https://doi.org/10.1088/1475-7516/2019/08/018}{\emph{JCAP} {\bfseries
  1908} (2019) 018} [\href{https://arxiv.org/abs/1906.04204}{{\ttfamily
  1906.04204}}].

\bibitem{Goldman:1989nd}
I.~Goldman and S.~Nussinov, \emph{{Weakly Interacting Massive Particles and
  Neutron Stars}}, \href{https://doi.org/10.1103/PhysRevD.40.3221}{\emph{Phys.
  Rev.} {\bfseries D40} (1989) 3221}.

\bibitem{Gould:1989gw}
A.~Gould, B.~T. Draine, R.~W. Romani and S.~Nussinov, \emph{{Neutron Stars:
  Graveyard of Charged Dark Matter}},
  \href{https://doi.org/10.1016/0370-2693(90)91745-W}{\emph{Phys. Lett.}
  {\bfseries B238} (1990) 337}.

\bibitem{Bertone:2007ae}
G.~Bertone and M.~Fairbairn, \emph{{Compact Stars as Dark Matter Probes}},
  \href{https://doi.org/10.1103/PhysRevD.77.043515}{\emph{Phys. Rev.}
  {\bfseries D77} (2008) 043515}
  [\href{https://arxiv.org/abs/0709.1485}{{\ttfamily 0709.1485}}].

\bibitem{deLavallaz:2010wp}
A.~de~Lavallaz and M.~Fairbairn, \emph{{Neutron Stars as Dark Matter Probes}},
  \href{https://doi.org/10.1103/PhysRevD.81.123521}{\emph{Phys. Rev.}
  {\bfseries D81} (2010) 123521}
  [\href{https://arxiv.org/abs/1004.0629}{{\ttfamily 1004.0629}}].

\bibitem{Kouvaris:2010jy}
C.~Kouvaris and P.~Tinyakov, \emph{{Constraining Asymmetric Dark Matter through
  observations of compact stars}},
  \href{https://doi.org/10.1103/PhysRevD.83.083512}{\emph{Phys. Rev.}
  {\bfseries D83} (2011) 083512}
  [\href{https://arxiv.org/abs/1012.2039}{{\ttfamily 1012.2039}}].

\bibitem{McDermott:2011jp}
S.~D. McDermott, H.-B. Yu and K.~M. Zurek, \emph{{Constraints on Scalar
  Asymmetric Dark Matter from Black Hole Formation in Neutron Stars}},
  \href{https://doi.org/10.1103/PhysRevD.85.023519}{\emph{Phys. Rev.}
  {\bfseries D85} (2012) 023519}
  [\href{https://arxiv.org/abs/1103.5472}{{\ttfamily 1103.5472}}].

\bibitem{Kouvaris:2011fi}
C.~Kouvaris and P.~Tinyakov, \emph{{Excluding Light Asymmetric Bosonic Dark
  Matter}}, \href{https://doi.org/10.1103/PhysRevLett.107.091301}{\emph{Phys.
  Rev. Lett.} {\bfseries 107} (2011) 091301}
  [\href{https://arxiv.org/abs/1104.0382}{{\ttfamily 1104.0382}}].

\bibitem{Guver:2012ba}
T.~Güver, A.~E. Erkoca, M.~Hall~Reno and I.~Sarcevic, \emph{{On the capture of
  dark matter by neutron stars}},
  \href{https://doi.org/10.1088/1475-7516/2014/05/013}{\emph{JCAP} {\bfseries
  1405} (2014) 013} [\href{https://arxiv.org/abs/1201.2400}{{\ttfamily
  1201.2400}}].

\bibitem{Kouvaris:2012dz}
C.~Kouvaris and P.~Tinyakov, \emph{{(Not)-constraining heavy asymmetric bosonic
  dark matter}}, \href{https://doi.org/10.1103/PhysRevD.87.123537}{\emph{Phys.
  Rev.} {\bfseries D87} (2013) 123537}
  [\href{https://arxiv.org/abs/1212.4075}{{\ttfamily 1212.4075}}].

\bibitem{Kouvaris:2013kra}
C.~Kouvaris and P.~Tinyakov, \emph{{Growth of Black Holes in the interior of
  Rotating Neutron Stars}},
  \href{https://doi.org/10.1103/PhysRevD.90.043512}{\emph{Phys. Rev.}
  {\bfseries D90} (2014) 043512}
  [\href{https://arxiv.org/abs/1312.3764}{{\ttfamily 1312.3764}}].

\bibitem{Bell:2013xk}
N.~F. Bell, A.~Melatos and K.~Petraki, \emph{{Realistic neutron star
  constraints on bosonic asymmetric dark matter}},
  \href{https://doi.org/10.1103/PhysRevD.87.123507}{\emph{Phys. Rev.}
  {\bfseries D87} (2013) 123507}
  [\href{https://arxiv.org/abs/1301.6811}{{\ttfamily 1301.6811}}].

\bibitem{Bramante:2013hn}
J.~Bramante, K.~Fukushima and J.~Kumar, \emph{{Constraints on bosonic dark
  matter from observation of old neutron stars}},
  \href{https://doi.org/10.1103/PhysRevD.87.055012}{\emph{Phys. Rev.}
  {\bfseries D87} (2013) 055012}
  [\href{https://arxiv.org/abs/1301.0036}{{\ttfamily 1301.0036}}].

\bibitem{Jamison:2013yya}
A.~O. Jamison, \emph{{Effects of gravitational confinement on bosonic
  asymmetric dark matter in stars}},
  \href{https://doi.org/10.1103/PhysRevD.88.035004}{\emph{Phys. Rev.}
  {\bfseries D88} (2013) 035004}
  [\href{https://arxiv.org/abs/1304.3773}{{\ttfamily 1304.3773}}].

\bibitem{Bramante:2013nma}
J.~Bramante, K.~Fukushima, J.~Kumar and E.~Stopnitzky, \emph{{Bounds on
  self-interacting fermion dark matter from observations of old neutron
  stars}}, \href{https://doi.org/10.1103/PhysRevD.89.015010}{\emph{Phys. Rev.}
  {\bfseries D89} (2014) 015010}
  [\href{https://arxiv.org/abs/1310.3509}{{\ttfamily 1310.3509}}].

\bibitem{Bramante:2014zca}
J.~Bramante and T.~Linden, \emph{{Detecting Dark Matter with Imploding Pulsars
  in the Galactic Center}},
  \href{https://doi.org/10.1103/PhysRevLett.113.191301}{\emph{Phys. Rev. Lett.}
  {\bfseries 113} (2014) 191301}
  [\href{https://arxiv.org/abs/1405.1031}{{\ttfamily 1405.1031}}].

\bibitem{Garani:2018kkd}
R.~Garani, Y.~Genolini and T.~Hambye, \emph{{New Analysis of Neutron Star
  Constraints on Asymmetric Dark Matter}},
  \href{https://doi.org/10.1088/1475-7516/2019/05/035}{\emph{JCAP} {\bfseries
  1905} (2019) 035} [\href{https://arxiv.org/abs/1812.08773}{{\ttfamily
  1812.08773}}].

\bibitem{Lin:2020zmm}
G.-L. Lin and Y.-H. Lin, \emph{{Analysis on the black hole formations inside
  old neutron stars by isospin-violating dark matter with self-interaction}},
  \href{https://doi.org/10.1088/1475-7516/2020/08/022}{\emph{JCAP} {\bfseries
  08} (2020) 022} [\href{https://arxiv.org/abs/2004.05312}{{\ttfamily
  2004.05312}}].

\bibitem{Bramante:2015cua}
J.~Bramante, \emph{{Dark matter ignition of type Ia supernovae}},
  \href{https://doi.org/10.1103/PhysRevLett.115.141301}{\emph{Phys. Rev. Lett.}
  {\bfseries 115} (2015) 141301}
  [\href{https://arxiv.org/abs/1505.07464}{{\ttfamily 1505.07464}}].

\bibitem{Acevedo:2019gre}
J.~F. Acevedo and J.~Bramante, \emph{{Supernovae Sparked By Dark Matter in
  White Dwarfs}},
  \href{https://doi.org/10.1103/PhysRevD.100.043020}{\emph{Phys. Rev.}
  {\bfseries D100} (2019) 043020}
  [\href{https://arxiv.org/abs/1904.11993}{{\ttfamily 1904.11993}}].

\bibitem{Graham:2018efk}
P.~W. Graham, R.~Janish, V.~Narayan, S.~Rajendran and P.~Riggins, \emph{{White
  Dwarfs as Dark Matter Detectors}},
  \href{https://doi.org/10.1103/PhysRevD.98.115027}{\emph{Phys. Rev.}
  {\bfseries D98} (2018) 115027}
  [\href{https://arxiv.org/abs/1805.07381}{{\ttfamily 1805.07381}}].

\bibitem{Janish:2019nkk}
R.~Janish, V.~Narayan and P.~Riggins, \emph{{Type Ia supernovae from dark
  matter core collapse}},
  \href{https://doi.org/10.1103/PhysRevD.100.035008}{\emph{Phys. Rev. D}
  {\bfseries 100} (2019) 035008}
  [\href{https://arxiv.org/abs/1905.00395}{{\ttfamily 1905.00395}}].

\bibitem{Kouvaris:2007ay}
C.~Kouvaris, \emph{{WIMP Annihilation and Cooling of Neutron Stars}},
  \href{https://doi.org/10.1103/PhysRevD.77.023006}{\emph{Phys. Rev.}
  {\bfseries D77} (2008) 023006}
  [\href{https://arxiv.org/abs/0708.2362}{{\ttfamily 0708.2362}}].

\bibitem{Kouvaris:2010vv}
C.~Kouvaris and P.~Tinyakov, \emph{{Can Neutron stars constrain Dark Matter?}},
  \href{https://doi.org/10.1103/PhysRevD.82.063531}{\emph{Phys. Rev.}
  {\bfseries D82} (2010) 063531}
  [\href{https://arxiv.org/abs/1004.0586}{{\ttfamily 1004.0586}}].

\bibitem{Baryakhtar:2017dbj}
M.~Baryakhtar, J.~Bramante, S.~W. Li, T.~Linden and N.~Raj, \emph{{Dark Kinetic
  Heating of Neutron Stars and An Infrared Window On WIMPs, SIMPs, and Pure
  Higgsinos}},
  \href{https://doi.org/10.1103/PhysRevLett.119.131801}{\emph{Phys. Rev. Lett.}
  {\bfseries 119} (2017) 131801}
  [\href{https://arxiv.org/abs/1704.01577}{{\ttfamily 1704.01577}}].

\bibitem{Bell:2018pkk}
N.~F. Bell, G.~Busoni and S.~Robles, \emph{{Heating up Neutron Stars with
  Inelastic Dark Matter}},
  \href{https://doi.org/10.1088/1475-7516/2018/09/018}{\emph{JCAP} {\bfseries
  1809} (2018) 018} [\href{https://arxiv.org/abs/1807.02840}{{\ttfamily
  1807.02840}}].

\bibitem{Chen:2018ohx}
C.-S. Chen and Y.-H. Lin, \emph{{Reheating neutron stars with the annihilation
  of self-interacting dark matter}},
  \href{https://doi.org/10.1007/JHEP08(2018)069}{\emph{JHEP} {\bfseries 08}
  (2018) 069} [\href{https://arxiv.org/abs/1804.03409}{{\ttfamily
  1804.03409}}].

\bibitem{Acevedo:2019agu}
J.~F. Acevedo, J.~Bramante, R.~K. Leane and N.~Raj, \emph{{Warming Nuclear
  Pasta with Dark Matter: Kinetic and Annihilation Heating of Neutron Star
  Crusts}}, \href{https://doi.org/10.1088/1475-7516/2020/03/038}{\emph{JCAP}
  {\bfseries 03} (2020) 038}
  [\href{https://arxiv.org/abs/1911.06334}{{\ttfamily 1911.06334}}].

\bibitem{McCullough:2010ai}
M.~McCullough and M.~Fairbairn, \emph{{Capture of Inelastic Dark Matter in
  White Dwarves}},
  \href{https://doi.org/10.1103/PhysRevD.81.083520}{\emph{Phys. Rev.}
  {\bfseries D81} (2010) 083520}
  [\href{https://arxiv.org/abs/1001.2737}{{\ttfamily 1001.2737}}].

\bibitem{Hooper:2010es}
D.~Hooper, D.~Spolyar, A.~Vallinotto and N.~Y. Gnedin, \emph{{Inelastic Dark
  Matter As An Efficient Fuel For Compact Stars}},
  \href{https://doi.org/10.1103/PhysRevD.81.103531}{\emph{Phys. Rev.}
  {\bfseries D81} (2010) 103531}
  [\href{https://arxiv.org/abs/1002.0005}{{\ttfamily 1002.0005}}].

\bibitem{Spergel:1999mh}
D.~N. Spergel and P.~J. Steinhardt, \emph{{Observational evidence for
  selfinteracting cold dark matter}},
  \href{https://doi.org/10.1103/PhysRevLett.84.3760}{\emph{Phys. Rev. Lett.}
  {\bfseries 84} (2000) 3760}
  [\href{https://arxiv.org/abs/astro-ph/9909386}{{\ttfamily
  astro-ph/9909386}}].

\bibitem{Dave:2000ar}
R.~Dave, D.~N. Spergel, P.~J. Steinhardt and B.~D. Wandelt, \emph{{Halo
  properties in cosmological simulations of selfinteracting cold dark matter}},
  \href{https://doi.org/10.1086/318417}{\emph{Astrophys. J.} {\bfseries 547}
  (2001) 574} [\href{https://arxiv.org/abs/astro-ph/0006218}{{\ttfamily
  astro-ph/0006218}}].

\bibitem{Loeb:2010gj}
A.~Loeb and N.~Weiner, \emph{{Cores in Dwarf Galaxies from Dark Matter with a
  Yukawa Potential}},
  \href{https://doi.org/10.1103/PhysRevLett.106.171302}{\emph{Phys. Rev. Lett.}
  {\bfseries 106} (2011) 171302}
  [\href{https://arxiv.org/abs/1011.6374}{{\ttfamily 1011.6374}}].

\bibitem{Aarssen:2012fx}
L.~G. van~den Aarssen, T.~Bringmann and C.~Pfrommer, \emph{{Is dark matter with
  long-range interactions a solution to all small-scale problems of $\Lambda$
  CDM cosmology?}},
  \href{https://doi.org/10.1103/PhysRevLett.109.231301}{\emph{Phys. Rev. Lett.}
  {\bfseries 109} (2012) 231301}
  [\href{https://arxiv.org/abs/1205.5809}{{\ttfamily 1205.5809}}].

\bibitem{Dasgupta:2013zpn}
B.~Dasgupta and J.~Kopp, \emph{{Cosmologically Safe eV-Scale Sterile Neutrinos
  and Improved Dark Matter Structure}},
  \href{https://doi.org/10.1103/PhysRevLett.112.031803}{\emph{Phys. Rev. Lett.}
  {\bfseries 112} (2014) 031803}
  [\href{https://arxiv.org/abs/1310.6337}{{\ttfamily 1310.6337}}].

\bibitem{Chu:2014lja}
X.~Chu and B.~Dasgupta, \emph{{Dark Radiation Alleviates Problems with Dark
  Matter Halos}},
  \href{https://doi.org/10.1103/PhysRevLett.113.161301}{\emph{Phys. Rev. Lett.}
  {\bfseries 113} (2014) 161301}
  [\href{https://arxiv.org/abs/1404.6127}{{\ttfamily 1404.6127}}].

\bibitem{Kaplinghat:2015aga}
M.~Kaplinghat, S.~Tulin and H.-B. Yu, \emph{{Dark Matter Halos as Particle
  Colliders: Unified Solution to Small-Scale Structure Puzzles from Dwarfs to
  Clusters}}, \href{https://doi.org/10.1103/PhysRevLett.116.041302}{\emph{Phys.
  Rev. Lett.} {\bfseries 116} (2016) 041302}
  [\href{https://arxiv.org/abs/1508.03339}{{\ttfamily 1508.03339}}].

\bibitem{Kamada:2016euw}
A.~Kamada, M.~Kaplinghat, A.~B. Pace and H.-B. Yu, \emph{{How the
  Self-Interacting Dark Matter Model Explains the Diverse Galactic Rotation
  Curves}}, \href{https://doi.org/10.1103/PhysRevLett.119.111102}{\emph{Phys.
  Rev. Lett.} {\bfseries 119} (2017) 111102}
  [\href{https://arxiv.org/abs/1611.02716}{{\ttfamily 1611.02716}}].

\bibitem{Tulin:2017ara}
S.~Tulin and H.-B. Yu, \emph{{Dark Matter Self-interactions and Small Scale
  Structure}}, \href{https://doi.org/10.1016/j.physrep.2017.11.004}{\emph{Phys.
  Rept.} {\bfseries 730} (2018) 1}
  [\href{https://arxiv.org/abs/1705.02358}{{\ttfamily 1705.02358}}].

\bibitem{Bullock:2017xww}
J.~S. Bullock and M.~Boylan-Kolchin, \emph{{Small-Scale Challenges to the
  $\Lambda$CDM Paradigm}},
  \href{https://doi.org/10.1146/annurev-astro-091916-055313}{\emph{Ann. Rev.
  Astron. Astrophys.} {\bfseries 55} (2017) 343}
  [\href{https://arxiv.org/abs/1707.04256}{{\ttfamily 1707.04256}}].

\bibitem{Dawson:2011kf}
W.~A. Dawson et~al., \emph{{Discovery of a Dissociative Galaxy Cluster Merger
  with Large Physical Separation}},
  \href{https://doi.org/10.1088/2041-8205/747/2/L42}{\emph{Astrophys. J.}
  {\bfseries 747} (2012) L42}
  [\href{https://arxiv.org/abs/1110.4391}{{\ttfamily 1110.4391}}].

\bibitem{Merten:2011wj}
J.~Merten et~al., \emph{{Creation of cosmic structure in the complex galaxy
  cluster merger Abell 2744}},
  \href{https://doi.org/10.1111/j.1365-2966.2011.19266.x}{\emph{Mon. Not. Roy.
  Astron. Soc.} {\bfseries 417} (2011) 333}
  [\href{https://arxiv.org/abs/1103.2772}{{\ttfamily 1103.2772}}].

\bibitem{Bradac:2008eu}
M.~Bradac, S.~W. Allen, T.~Treu, H.~Ebeling, R.~Massey, R.~G. Morris et~al.,
  \emph{{Revealing the properties of dark matter in the merging cluster
  MACSJ0025.4-1222}}, \href{https://doi.org/10.1086/591246}{\emph{Astrophys.
  J.} {\bfseries 687} (2008) 959}
  [\href{https://arxiv.org/abs/0806.2320}{{\ttfamily 0806.2320}}].

\bibitem{Randall:2007ph}
S.~W. Randall, M.~Markevitch, D.~Clowe, A.~H. Gonzalez and M.~Bradac,
  \emph{{Constraints on the Self-Interaction Cross-Section of Dark Matter from
  Numerical Simulations of the Merging Galaxy Cluster 1E 0657-56}},
  \href{https://doi.org/10.1086/587859}{\emph{Astrophys. J.} {\bfseries 679}
  (2008) 1173} [\href{https://arxiv.org/abs/0704.0261}{{\ttfamily 0704.0261}}].

\bibitem{Robertson:2016xjh}
A.~Robertson, R.~Massey and V.~Eke, \emph{{What does the Bullet Cluster tell us
  about self-interacting dark matter?}},
  \href{https://doi.org/10.1093/mnras/stw2670}{\emph{Mon. Not. Roy. Astron.
  Soc.} {\bfseries 465} (2017) 569}
  [\href{https://arxiv.org/abs/1605.04307}{{\ttfamily 1605.04307}}].

\bibitem{Zentner:2009is}
A.~R. Zentner, \emph{{High-Energy Neutrinos From Dark Matter Particle
  Self-Capture Within the Sun}},
  \href{https://doi.org/10.1103/PhysRevD.80.063501}{\emph{Phys. Rev.}
  {\bfseries D80} (2009) 063501}
  [\href{https://arxiv.org/abs/0907.3448}{{\ttfamily 0907.3448}}].

\bibitem{Kouvaris:2011gb}
C.~Kouvaris, \emph{{Limits on Self-Interacting Dark Matter}},
  \href{https://doi.org/10.1103/PhysRevLett.108.191301}{\emph{Phys. Rev. Lett.}
  {\bfseries 108} (2012) 191301}
  [\href{https://arxiv.org/abs/1111.4364}{{\ttfamily 1111.4364}}].

\bibitem{Fan:2013bea}
J.~Fan, A.~Katz and J.~Shelton, \emph{{Direct and indirect detection of
  dissipative dark matter}},
  \href{https://doi.org/10.1088/1475-7516/2014/06/059}{\emph{JCAP} {\bfseries
  1406} (2014) 059} [\href{https://arxiv.org/abs/1312.1336}{{\ttfamily
  1312.1336}}].

\bibitem{Chen:2015uha}
J.~Chen, Z.-L. Liang, Y.-L. Wu and Y.-F. Zhou, \emph{{Long-range
  self-interacting dark matter in the Sun}},
  \href{https://doi.org/10.1088/1475-7516/2015/12/021}{\emph{JCAP} {\bfseries
  1512} (2015) 021} [\href{https://arxiv.org/abs/1505.04031}{{\ttfamily
  1505.04031}}].

\bibitem{Chen:2015bwa}
C.-S. Chen, G.-L. Lin and Y.-H. Lin, \emph{{Complementary Test of the Dark
  Matter Self-Interaction by Direct and Indirect Detections}},
  \href{https://doi.org/10.1088/1475-7516/2016/01/013}{\emph{JCAP} {\bfseries
  1601} (2016) 013} [\href{https://arxiv.org/abs/1505.03781}{{\ttfamily
  1505.03781}}].

\bibitem{Feng:2016ijc}
J.~L. Feng, J.~Smolinsky and P.~Tanedo, \emph{{Detecting dark matter through
  dark photons from the Sun: Charged particle signatures}},
  \href{https://doi.org/10.1103/PhysRevD.93.115036,
  10.1103/PhysRevD.96.099903}{\emph{Phys. Rev.} {\bfseries D93} (2016) 115036}
  [\href{https://arxiv.org/abs/1602.01465}{{\ttfamily 1602.01465}}].

\bibitem{Gaidau:2018yws}
C.~Gaidau and J.~Shelton, \emph{{A Solar System Test of Self-Interacting Dark
  Matter}}, \href{https://doi.org/10.1088/1475-7516/2019/06/022}{\emph{JCAP}
  {\bfseries 1906} (2019) 022}
  [\href{https://arxiv.org/abs/1811.00557}{{\ttfamily 1811.00557}}].

\bibitem{Ilie:2020vec}
C.~Ilie, J.~Pilawa and S.~Zhang, \emph{{Comment on
  \textquotedblleft{}Multiscatter stellar capture of dark
  matter\textquotedblright{}}},
  \href{https://doi.org/10.1103/PhysRevD.102.048301}{\emph{Phys. Rev. D}
  {\bfseries 102} (2020) 048301}
  [\href{https://arxiv.org/abs/2005.05946}{{\ttfamily 2005.05946}}].

\bibitem{Joglekar:2020liw}
A.~Joglekar, N.~Raj, P.~Tanedo and H.-B. Yu, \emph{{Kinetic Heating from
  Contact Interactions with Relativistic Targets: Electrons Capture Dark Matter
  in Neutron Stars}},  \href{https://arxiv.org/abs/2004.09539}{{\ttfamily
  2004.09539}}.

\bibitem{Bell:2020jou}
N.~F. Bell, G.~Busoni, S.~Robles and M.~Virgato, \emph{{Improved Treatment of
  Dark Matter Capture in Neutron Stars}},
  \href{https://doi.org/10.1088/1475-7516/2020/09/028}{\emph{JCAP} {\bfseries
  09} (2020) 028} [\href{https://arxiv.org/abs/2004.14888}{{\ttfamily
  2004.14888}}].

\bibitem{Guillot:2019ugf}
S.~Guillot, G.~Pavlov, C.~Reyes, A.~Reisenegger, L.~Rodriguez, B.~Rangelov
  et~al., \emph{{Hubble Space Telescope Nondetection of PSR J2144--3933: The
  Coldest Known Neutron Star}},
  \href{https://doi.org/10.3847/1538-4357/ab0f38}{\emph{Astrophys. J.}
  {\bfseries 874} (2019) 175}
  [\href{https://arxiv.org/abs/1901.07998}{{\ttfamily 1901.07998}}].

\bibitem{Gardner:2006ky}
J.~P. Gardner et~al., \emph{{The James Webb Space Telescope}},
  \href{https://doi.org/10.1007/s11214-006-8315-7}{\emph{Space Sci. Rev.}
  {\bfseries 123} (2006) 485}
  [\href{https://arxiv.org/abs/astro-ph/0606175}{{\ttfamily
  astro-ph/0606175}}].

\bibitem{Crampton:2008gx}
D.~Crampton, L.~Simard and D.~Silva, \emph{{TMT Science and Instruments}},
  \href{https://doi.org/10.1007/978-1-4020-9190-2\_47}{\emph{Astrophys. Space
  Sci. Proc.} (2009) 279} [\href{https://arxiv.org/abs/0801.3634}{{\ttfamily
  0801.3634}}].

\bibitem{Maiolino:2013bsa}
R.~Maiolino et~al., \emph{{A Community Science Case for E-ELT HIRES}},
  \href{https://arxiv.org/abs/1310.3163}{{\ttfamily 1310.3163}}.

\bibitem{Hawking:1974rv}
S.~Hawking, \emph{{Black hole explosions}},
  \href{https://doi.org/10.1038/248030a0}{\emph{Nature} {\bfseries 248} (1974)
  30}.

\bibitem{Hawking:1974sw}
S.~Hawking, \emph{{Particle Creation by Black Holes}},
  \href{https://doi.org/10.1007/BF02345020}{\emph{Commun. Math. Phys.}
  {\bfseries 43} (1975) 199}.

\bibitem{MacGibbon:1990zk}
J.~MacGibbon and B.~Webber, \emph{{Quark and gluon jet emission from primordial
  black holes: The instantaneous spectra}},
  \href{https://doi.org/10.1103/PhysRevD.41.3052}{\emph{Phys. Rev. D}
  {\bfseries 41} (1990) 3052}.

\bibitem{MacGibbon:1991tj}
J.~H. MacGibbon, \emph{{Quark and gluon jet emission from primordial black
  holes. 2. The Lifetime emission}},
  \href{https://doi.org/10.1103/PhysRevD.44.376}{\emph{Phys. Rev. D} {\bfseries
  44} (1991) 376}.

\bibitem{Dasgupta:2019cae}
B.~Dasgupta, R.~Laha and A.~Ray, \emph{{Neutrino and positron constraints on
  spinning primordial black hole dark matter}},
  \href{https://doi.org/10.1103/PhysRevLett.125.101101}{\emph{Phys. Rev. Lett.}
  {\bfseries 125} (2020) 101101}
  [\href{https://arxiv.org/abs/1912.01014}{{\ttfamily 1912.01014}}].

\bibitem{Carr:2020gox}
B.~Carr, K.~Kohri, Y.~Sendouda and J.~Yokoyama, \emph{{Constraints on
  Primordial Black Holes}},  \href{https://arxiv.org/abs/2002.12778}{{\ttfamily
  2002.12778}}.

\bibitem{Garani:2019fpa}
R.~Garani and J.~Heeck, \emph{{Dark matter interactions with muons in neutron
  stars}}, \href{https://doi.org/10.1103/PhysRevD.100.035039}{\emph{Phys. Rev.}
  {\bfseries D100} (2019) 035039}
  [\href{https://arxiv.org/abs/1906.10145}{{\ttfamily 1906.10145}}].

\bibitem{Joglekar:2019vzy}
A.~Joglekar, N.~Raj, P.~Tanedo and H.-B. Yu, \emph{{Relativistic capture of
  dark matter by electrons in neutron stars}},
  \href{https://doi.org/10.1016/j.physletb.2020.135767}{\emph{Phys. Lett.}
  {\bfseries B} (2020) 135767}
  [\href{https://arxiv.org/abs/1911.13293}{{\ttfamily 1911.13293}}].

\bibitem{Bell:2019pyc}
N.~F. Bell, G.~Busoni and S.~Robles, \emph{{Capture of Leptophilic Dark Matter
  in Neutron Stars}},
  \href{https://doi.org/10.1088/1475-7516/2019/06/054}{\emph{JCAP} {\bfseries
  06} (2019) 054} [\href{https://arxiv.org/abs/1904.09803}{{\ttfamily
  1904.09803}}].

\end{thebibliography}\endgroup

\end{document}